\documentclass[aps,prd,superscriptaddress,amsmath,amssymb,twocolumn,fleqn,nofootinbib]{revtex4-1} 

\usepackage{xcolor}
\usepackage{graphicx}
\usepackage{colordvi}
\usepackage{mathtools}
\usepackage{empheq}
\usepackage{bm}%

\usepackage{floatrow}
\usepackage{braket}
\usepackage{soul}

\usepackage{amssymb}
\usepackage{amsmath}

\usepackage{hyperref}
\def\be{\begin{equation}}
\def\ee{\end{equation}}
\def\ba{\begin{eqnarray}}
\def\ea{\end{eqnarray}}

\newcommand{\dd}{\mathrm{d}}

\begin{document}

\title{Early modified gravity in light of the $H_0$ tension and LSS data}

\author{Matteo Braglia}\email{matteo.braglia2@unibo.it}
\affiliation{Dipartimento di Fisica e Astronomia, Alma Mater Studiorum
Universit\`a di Bologna, \\
Via Gobetti, 93/2, I-40129 Bologna, Italy}
\affiliation{INAF/OAS Bologna, via Gobetti 101, I-40129 Bologna, Italy}
\affiliation{INFN, Sezione di Bologna, Via Berti Pichat 6/2, I-40127 Bologna, Italy}
\author{Mario Ballardini}\email{mario.ballardini@inaf.it}
\affiliation{Dipartimento di Fisica e Astronomia, Alma Mater Studiorum
	Universit\`a di Bologna, \\
	Via Gobetti, 93/2, I-40129 Bologna, Italy}
\affiliation{INAF/OAS Bologna, via Gobetti 101, I-40129 Bologna, Italy}
\affiliation{INFN, Sezione di Bologna, Via Berti Pichat 6/2, I-40127 Bologna, Italy}

\author{Fabio Finelli}\email{fabio.finelli@inaf.it}
\affiliation{INAF/OAS Bologna, via Gobetti 101, I-40129 Bologna, Italy}
\affiliation{INFN, Sezione di Bologna, Via Berti Pichat 6/2, I-40127 Bologna, Italy}

\author{Kazuya Koyama}\email{kazuya.koyama@port.ac.uk}
\affiliation{Institute of Cosmology and Gravitation, University of Portsmouth, Dennis Sciama Building, Portsmouth PO1 3FX, United Kingdom}

\date{\today}
\begin{abstract}
We present a model of Early Modified Gravity (EMG) consisting in a scalar field $\sigma$ with a non-minimal coupling to the Ricci curvature of the type $M^2_{\rm pl}+\xi \sigma^2$ plus a cosmological constant and a small effective mass and demonstrate its ability to alleviate the $H_0$ tension while providing a good fit to Cosmic Microwave Background (CMB) anisotropies and Baryon Acoustic Oscillations (BAO) data. In this model the scalar field, frozen deep in the radiation era, 
grows around the redshift of matter-radiation equality because of the coupling to non-relativistic matter. 
The small effective mass, which we consider here as induced by a quartic potential,  
then damps the scalar field into coherent oscillations around its minimum at $\sigma=0$, leading to a weaker gravitational strength at early times and naturally recovering the consistency with laboratory and Solar System tests of gravity.  
We analyze the capability of EMG with positive $\xi$ to fit current cosmological observations and compare our results to the case without an effective mass and to the popular early dark energy models with $\xi=0$. We show that EMG with a quartic coupling of the order of $\lambda\sim\mathcal{O}({\rm eV}^4/M_{\rm pl}^4)$
can substantially alleviate the $H_0$ tension also when the full shape of the matter power spectrum is included in the fit in addition to CMB  and Supernovae (SN) data. 
\end{abstract}

      \maketitle
      
\section{Introduction}

The long-standing success of the cosmological $\Lambda$CDM model has been challenged in the recent years by the growing discrepancy between direct measurements of the Hubble constant $H_0$ and its inference from CMB anisotropies data \cite{Verde:2019ivm}.  The most recent measurements range from $H_0 = (67.36 \pm 0.54)$ km s$^{-1}$Mpc$^{-1}$ for $\Lambda$CDM and Planck 2018 data release \cite{Aghanim:2018eyx} and
 $H_0 = (73.5 \pm 1.4)$  km s$^{-1}$Mpc$^{-1}$ \cite{Reid:2019tiq} for SH0ES, showing a 4.1$\sigma$ tension on the $H_0$ parameter. However, the tension is not restricted to these two data sets. With the recent progress, it is now clear that, rather than being only between Planck and SH0ES, the $H_0$ tension is 
 in general between indirect, or early time, measurements obtained by inferring $H_0$ assuming a model (usually $\Lambda$CDM) and analyzing it with cosmological data such as the CMB \cite{Aghanim:2018eyx} or the combination of clustering and weak lensing data with BAO and Big Bang Nucleosynthesis ones \cite{Abbott:2017smn}, and direct, or late time,  measurements of $H_0$, which are instead model independent. 
 A number of $H_0$ probes belonging to the latter class are in tension with estimates from CMB up to $\sim 4 \sigma$ level, see \cite{Verde:2019ivm} for a review. 
 Another independent determination of $H_0$, important for this paper, is obtained with the strong-lensing time delay by the H0LiCOW team  \cite{Wong:2019kwg}, i.e. $H_0 = \left(73.3_{-1.8}^{+1.7}\right)$  km s$^{-1}$Mpc$^{-1}$, which is in a 3.2$\sigma$ tension with CMB (see however \cite{Birrer:2020tax} and \cite{Krishnan:2020obg} for implications on Early time solutions of the $H_0$ tension). 
 By combining SH0ES and H0LiCOW measurements the estimate  $H_0 = \left(73.4\pm1.1\right)$ km s$^{-1}$Mpc$^{-1}$ is obtained, raising the tension with CMB to the 4.9$\sigma$ level.
 Given the relevance of this tension, 
several groups have investigated whether it might be due to unaccounted effects such as uncertainties in calibration \cite{Efstathiou:2013via,Freedman:2019jwv,Freedman:2020dne,Yuan:2019npk} or in the luminosity functions of SNIa \cite{Efstathiou:2013via,Rigault:2014kaa,Rigault:2018ffm,Freedman:2019jwv,Freedman:2020dne,Yuan:2019npk,Efstathiou:2020wxn}. 

 Although unaccounted systematic effects might alter its statistical significance, 
these discrepant determinations of $H_0$ spark interest
towards new physics beyond $\Lambda$CDM \cite{Mortsell:2018mfj}. This point of view has stimulated the proposal of a wealth of physical mechanisms leading to a large $H_0$ through modifications of both the early \cite{DiValentino:2017oaw,DEramo:2018vss,Poulin:2018zxs,Kreisch:2019yzn,Blinov:2019gcj,Pandey:2019plg} and the late time 
\cite{DiValentino:2016hlg,Keeley:2019esp,Vagnozzi:2019ezj,DiValentino:2019ffd,Benevento:2020fev,Alestas:2020mvb,DiValentino:2020kha,Calderon:2020hoc} expansion history of the Universe. The former ones, however, seem to be preferred over since reduce the value of the comoving size of the sound horizon at baryon drag $r_s$, 
without spoiling the fit to CMB and BAO data \cite{Bernal:2016gxb,Aylor:2018drw,Knox:2019rjx}. 
Two well studied frameworks to modify the early time dynamics of the Universe and inject the required energy into the cosmic fluid to lower $r_s$ with respect to the $\Lambda$CDM one are modified gravity  (MG)
\cite{Umilta:2015cta,Ballardini:2016cvy,Nunes:2018xbm,Lin:2018nxe,Rossi:2019lgt,Sola:2019jek,Zumalacarregui:2020cjh,Wang:2020zfv,Ballesteros:2020sik,Braglia:2020iik,Ballardini:2020iws,Sola:2020lba,Joudaki:2020shz}
and Early Dark Energy (EDE) models
\cite{Poulin:2018dzj,Poulin:2018cxd,Agrawal:2019lmo,Kaloper:2019lpl,Alexander:2019rsc,Niedermann:2019olb,Berghaus:2019cls,Sakstein:2019fmf,Lin:2019qug,Smith:2019ihp,Ye:2020btb,Braglia:2020bym,Niedermann:2020dwg,Gonzalez:2020fdy,Niedermann:2020qbw,Lin:2020jcb,CarrilloGonzalez:2020oac,Das:2020wfe}\footnote{  See also Refs.~\cite{Sahni:2014ooa,Shafieloo:2018gin,LHuillier:2019imn} for ways to constrain
EDE or more in general Dark Energy models with a time varying equation of state 
based on the reconstruction of the Universe expansion from the density growth factor redshift dependence.}.

In this paper, we extend the model with a scalar field on-minimally coupled to the Ricci scalar of the form
$F(\sigma)=M_\textup{pl}^2+\xi\sigma^2$ in presence of a cosmological constant $\Lambda$ \cite{Ballesteros:2020sik,Braglia:2020iik}, by providing it with a small effective mass. For the sake of simplicity we consider a quartic potential, i.e.  $V(\sigma)=\lambda\,\sigma^4/4$, as effective mass: in this Early Modified Gravity (EMG) model,
the scalar field starts to move around the redshift of matter-radiation equality driven by the
coupling to non-relativistic matter, and then rolls faster when the effective mass become larger than the Hubble parameter
and ends in a regime of coherent oscillations around the minimum of the potential. The choice of a quartic potential is dictated by the fact that coherent oscillations of $\sigma$ are in conformal time and therefore tractable by an Einstein-Boltzmann code, without ad-hoc modifications, see e.g. Ref.~\cite{Hlozek:2016lzm}. Thanks to the fast rolling of $\sigma$ towards the bottom of the potential, the tight constraints on $G_{\rm eff}$ from laboratory
experiments and Solar System measurements on post-Newtonian parameters are automatically satisfied by the small cosmological values of $\sigma$
within the EMG model, as it happens in the range of $\xi < 0$ in the massless case where $\sigma$ is decreased just by coupling to non-relativistic matter \cite{Rossi:2019lgt,Ballesteros:2020sik,Braglia:2020iik}.
The small effective mass and the consequent naturally achieved consistency of cosmology with laboratory and Solar System constraints are particularly important
for positive values of the coupling, since $\sigma$ would grow for $\xi >0$ for $\lambda=0$, and therefore we mainly focus on this range in this paper.

 In our EMG model, we consider the two possible dimensionless couplings for a cosmological scalar field, which rule the coupling to the Ricci scalar ($\xi$) and 
its self-interaction ($\lambda$), respectively. Note that our model differs from previously introduced ones also named Early Modified Gravity \cite{Brax:2013fda,Lima:2016npg,Pettorino:2014bka}. 

Another interesting feature of this EMG is that the effective Newtonian constant $G_{\rm eff}$ grows with time, as opposed to nearly massless models \cite{Rossi:2019lgt,Ballesteros:2020sik,Braglia:2020iik,Ballardini:2020iws}, implying a weaker gravity at early times. As we show in this paper, such an effect implies different predictions on Large Scale Structure (LSS) observables that can help disentangle EMG and EDE. The latter models have indeed been recently claimed not to be able to solve the $H_0$ tension when LSS data are included in the analysis \cite{Hill:2020osr,Ivanov:2020ril,DAmico:2020ods}. As we show in this paper, the suppression of the matter power spectrum induced by the positive coupling helps us obtain a value for $H_0$  larger than EDE with $\xi=0$, thanks to a better fit to LSS data.

Our paper is organized as follows. We introduce our model and describe in details its background evolution, as well as its imprints on CMB and LSS observables in Sec.~\ref{sec:model}. We describe the dataset and the methodology used in our MCMC exploration in Sec.~\ref{sec:data},  present our results in Sec.~\ref{sec:results} and compare them with existing works on the EDE and NMC models in Sections~\ref{sec:1par} and \ref{sec:conf}, before concluding in Sec.~\ref{sec:conclusion}. We collect the tables with the results of our MCMC analysis in Appendix~\ref{sec:appendix}.

\section{The model} \label{sec:model}
The model we consider is described by the following action:
\begin{align}
\label{eq:model}
    S =& \int \dd^{4}x \sqrt{-g} \left[ \frac{F(\sigma)}{2}R 
    - \frac{g^{\mu\nu}}{2} \partial_\mu \sigma \partial_\nu \sigma -\Lambda- V(\sigma)\right]\notag\\& + S_m \,,
\end{align}
where $F(\sigma) = M_{pl}^2+\xi\sigma^2$, $R$ is the Ricci scalar, and $S_m$ is the
action for matter fields. In the following, we consider a quartic potential for the scalar field of the form $V(\sigma)=\lambda\,\sigma^4/4$, where $\lambda$ is a dimensionless constant. 
With these conventions, our model reduces to the NMC model  considered in Ref.~\cite{Braglia:2020iik} for $\lambda=0$ and to the Rock 'n' Roll (RnR) model of Ref.~\cite{Agrawal:2019lmo} for $\xi=0$. Since the latter is an example of Early Dark Energy models, we refer to it as EDE in the following and use the acronym EMG for the general case with $\xi\neq0$.

The  Friedmann  and Klein-Gordon equations are given by:
\begin{subequations}\label{eq:eoms}
	\begin{align}
	3 F H^2 \ =& \ \rho \: + \: \frac{\dot{\sigma}^2}{2} \: + \: \Lambda \:+ \: V \: - \: 3\dot{F}H  \equiv\ \rho \: + \: \rho_\sigma \;, 
	\end{align}
	\begin{align}
	\label{eq:KG}
 	\ddot{\sigma} \: + \: 3H\dot{\sigma} =& \ \frac{F_{\sigma}}{2F + 3F^2_{\sigma}}\Big[\rho \: - \: 3p \: + \: 4\Lambda \:+ \: 4 V\:\notag\\&- \: 2\frac{F\,V_\sigma}{F_\sigma} \:  - \: \big(1 \: + \: 3F_{\sigma\sigma}\big)\dot{\sigma}^2 \: \Big] \;, 
	\end{align}
\end{subequations}
where $\rho\equiv\rho_m+\rho_r$ ($p\equiv p_r$) denotes the sum of the matter and radiation energy density (pressure) and a subscript $\sigma$ denotes the derivative with respect to the scalar field $\sigma$. 
For theories described by the action \ref{eq:model}, it is useful to define an effective dark energy density as follows \cite{Boisseau:2000pr,Gannouji:2006jm}:
\begin{equation}
\rho_{\rm DE}=\frac{F_0}{F}\rho_\sigma+(\rho_m+\rho_r)\left(\frac{F_0}{F}-1\right),
\end{equation} where the subscript $0$ denotes that a quantity is evaluated at $z=0$.
The energy fraction of the scalar field is simply given by $\Omega_\sigma\equiv\rho_{\rm DE}/3 H^2 F_0$.

\begin{figure}
		\begin{center} 
			\includegraphics[width=.49\columnwidth]{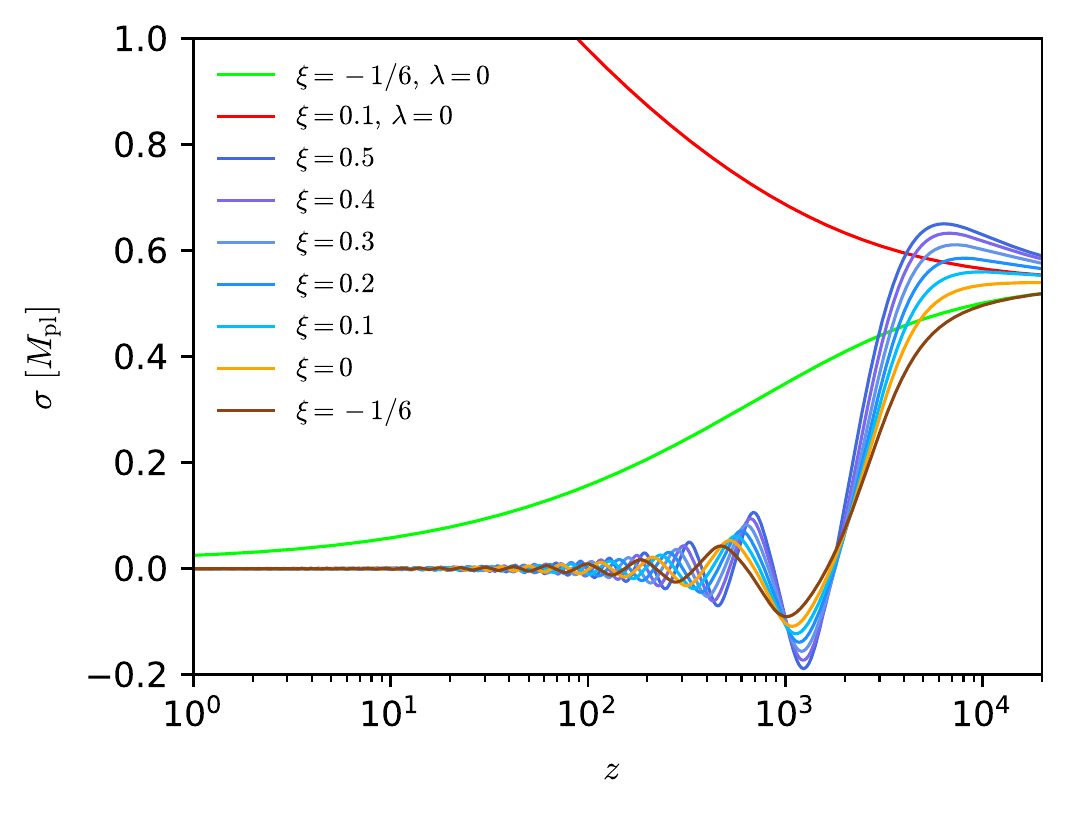}
			\includegraphics[width=.49\columnwidth]{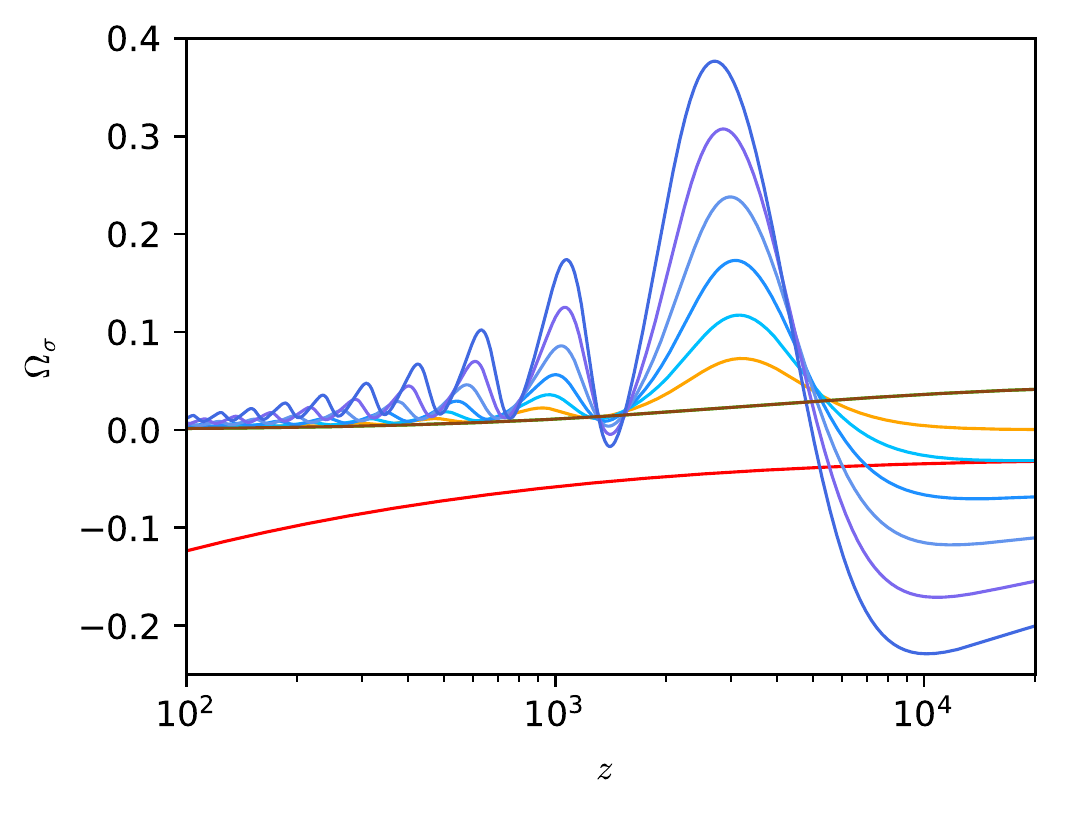}	
			\includegraphics[width=.49\columnwidth]{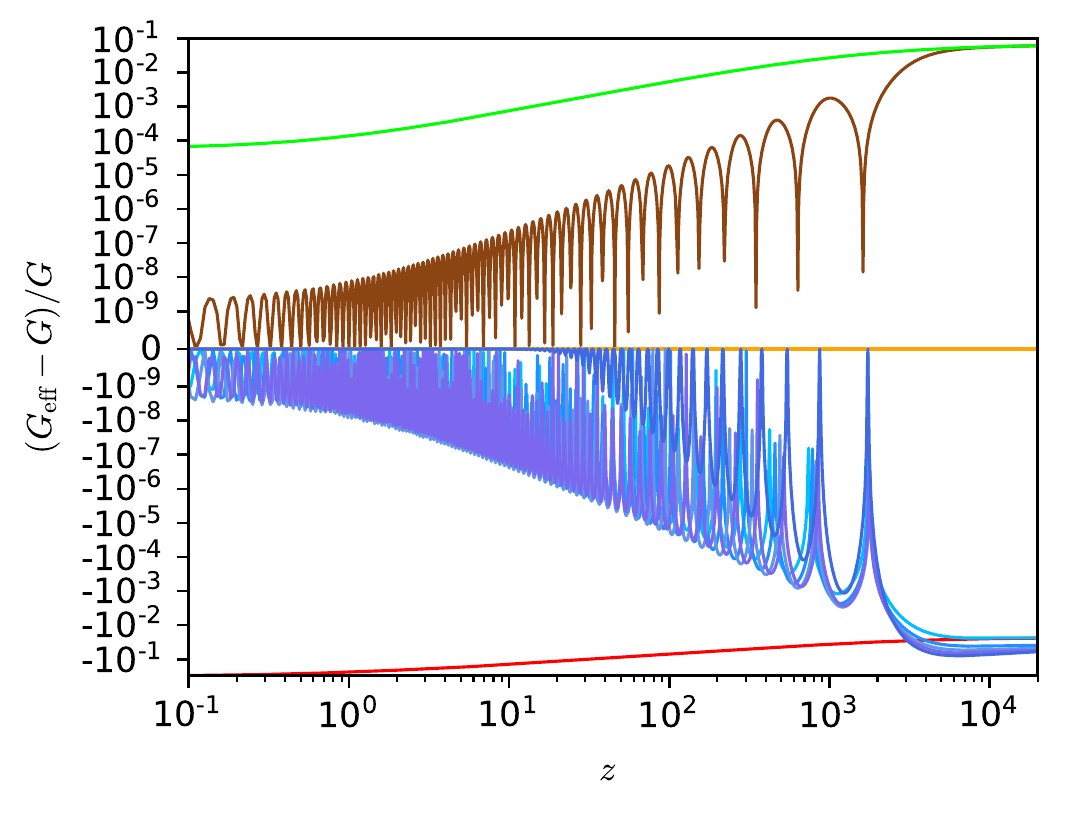}
			\includegraphics[width=.49\columnwidth]{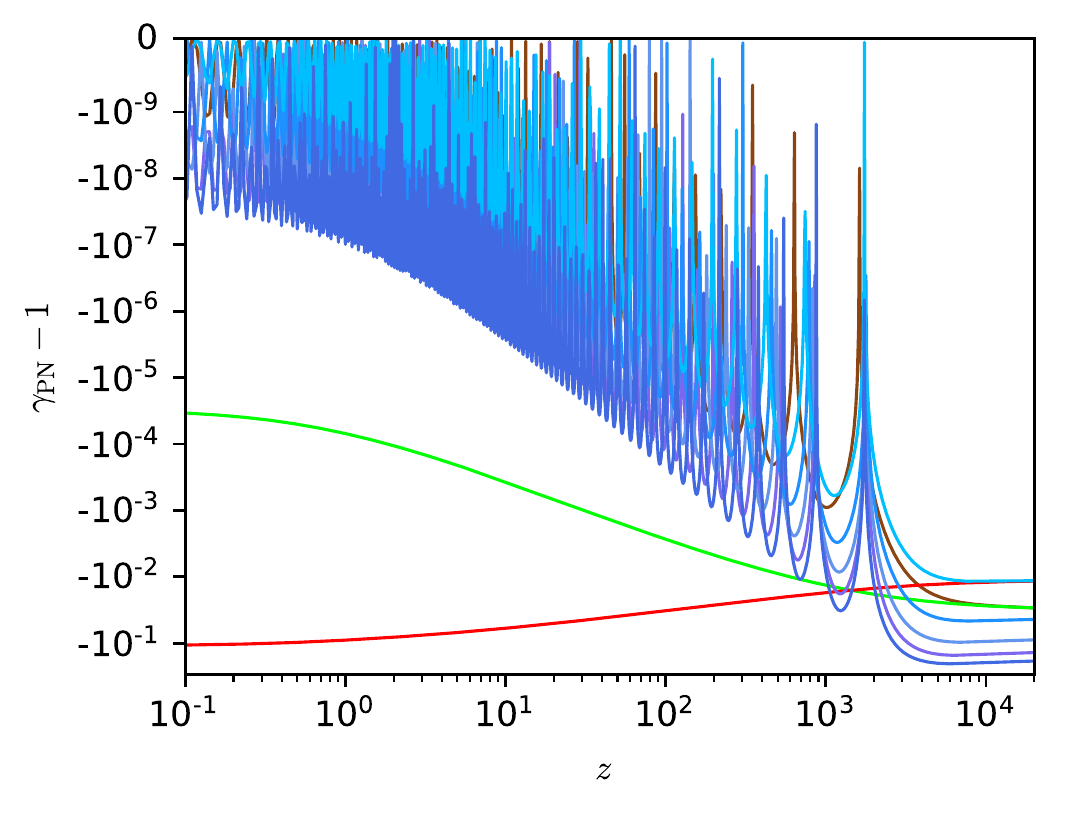}	
		\end{center}
		\caption{\label{fig:Background} 
		[Top] We plot the evolution of the scalar field (left) and the energy injection $\Omega_\sigma$ defined in the main text. [Bottom] We plot the  evolution of the variation of the effective Newton constant $(G_{\rm eff}-G)/G$  (left) and of the post-Newtonian parameter $\gamma_{\rm PN}-1$ (right). The model parameters used in the plot are $\sigma_i= 0.54\, M_{\rm pl}$ and $V_0=2$ and we vary the value of the non-minimal coupling $\xi$ according to the legend in the top-left panel. For a comparison, we also plot two examples for $\lambda=0$ in Eq.~\eqref{eq:lambda} and $\xi=0.1$ ($-1/6$) in red (green).  }
	\end{figure}

The coupling between gravity and the scalar degree of freedom induces a time varying Newton's 
gravitational constant $G_N$, which is given by $G_N = 1/(8\pi F)$. This quantity  is usually 
named the cosmological Newton's gravitational constant, as opposed to the one that is actually 
measured in laboratory Cavendish-type experiments which, for a nearly massless scalar tensor theory of gravity, is rather given by \cite{Boisseau:2000pr}:
\begin{equation}
\label{eq:Geff}
    G_{\rm eff} = \frac{1}{8\pi F}\frac{2F+4F^2_{\sigma}}{2F+3F^2_{\sigma}} \,.
\end{equation}
Note that, strictly speaking, given the non-vanishing potential for the scalar field $V(\sigma)=\lambda\,\sigma^4/4$, we would have scale-dependent fifth forces corrections in $G_{\rm eff}$ that  are proportional to $V_\sigma$ and $V_{\sigma\sigma}$ (see e.g. Refs.~\cite{Koyama:2015vza,Alby:2017dzl}). However, since $V_\sigma\simeq V_{\sigma\sigma}\simeq 0$ at lat times for the models considered in this paper (see also Fig.~\ref{fig:Background}), such scale dependent corrections vanish and so we will use Eq.~\eqref{eq:Geff} throughout this paper. 

\onecolumngrid

\twocolumngrid

The deviations from general relativity (GR) can also be parameterized by means of the so-called Post-Newtonian (PN) parameters \cite{Will:2014kxa}, which are given within NMC by the following equations \cite{Boisseau:2000pr}: 
\begin{align}
\label{eqn:gammaPN}
\gamma_{\rm PN}&=1-\frac{F_{\sigma}^{2}}{F+2F_{\sigma}^{2}},\\
\label{eqn:betaPN}
\beta_{\rm PN}&=1+\frac{F F_{\sigma}}{8F+12F_{\sigma}^{2}}\frac{\dd\gamma_{\rm PN}}{\dd\sigma}.
\end{align}
 Note that $\gamma_{\rm PN}<1$ in our models.
Solar-system experiments agree with GR predictions, for which both $\gamma_{\rm PN}$ and 
$\beta_{\rm PN}$  are identically equal to unity, at a very precise level. Measurements of 
the perihelion shift of Mercury constrain $\beta_{\rm PN}-1 = (4.1\pm7.8) \times 10^{-5}$ at 68\% CL 
\cite{Will:2014kxa} and Shapiro time delay constrains 
$\gamma_{\rm PN}-1 = (2.1 \pm 2.3) \times 10^{-5}$ at 68\% CL \cite{Bertotti:2003rm}. As we will see below, such limits are automatically satisfied in our model.

\subsection{Background evolution}

The evolution of relevant background quantities is shown in Fig.~\ref{fig:Background}. For our comparison, we consider the bestfit cosmological parameters given in Table~3 of Ref.~\cite{Agrawal:2019lmo}, that is 
\begin{align}
    &\theta_s=1.0417,\,\,\,\, 100\, \omega_b=2.264,\, \,\,\,\,   \omega_c=0.1267,\,\,\,\,\notag\\&\tau_{\rm reio}=0.081,\,\,\,\,\ln10^{10}A_s=3.105,\,\,\,\,n_s=0.981,\,\,\,\,\notag\\& \sigma_i [M_{\rm pl}] =0.54,\,\,\,\,\,V_0=2\label{eq:paramsRnR}
\end{align} for EMG, where $\sigma_i$ is the initial condition on the scalar field and  for which we vary the non-minimal coupling $\xi$ according to the legend in the figures, and
\begin{align}
    \theta_s=1.0422,\,\,\,\, 100\, \omega_b=2.236,\, \,\,\,   \omega_c=0.1177,\,\,\,\,\notag\\\tau_{\rm reio}=0.077,\,\,\,\,\ln10^{10}A_s=3.080,\,\,\,\,n_s=0.969\label{eq:paramsLCDM}
\end{align}
for the $\Lambda$CDM model to which we compare our results. 
Note that the constant $V_0$ is related to $\lambda$ by\footnote{$3.516\times10^{109}$ is the \emph{numerical} value of $M_{\rm pl}^4$ in ${\rm eV}^4$} 
\begin{equation}
\label{eq:lambda}
    \lambda=10^{2 V_0}/(3.516\times10^{109}).
\end{equation}
We stress that these values are only used to build our intuition and will be superseded the cosmological parameter estimation that we present in Sec.~\ref{sec:results}. As can be seen from the top-left panel, 
the addition of the effective mass makes EMG more similar to EDE models with respect to nearly massless NMC models \cite{Rossi:2019lgt,Ballesteros:2020sik,Braglia:2020iik}.
Indeed, $\sigma$ starts frozen deep in the radiation era and, when its effective mass becomes larger than the Hubble flow, eventually rolls down the potential and starts oscillating around its effective minimum located at $\sigma=0$. It is clear from Fig.~\ref{fig:Background}, that the corrections to the effective mass of the scalar field induced by the non-minimal coupling $F(\sigma)$ modify the dynamics of $\sigma$, which, for $\xi\geq0$, experiences a temporary growth before falling down the potential.  Because of this initial growth, the oscillations around $\sigma=0$ have a visibly larger amplitude and their phase is slightly shifted compared to the case with $\xi=0$.  

The importance of such a modification to the dynamics for $\xi=0$ can be understood by looking at the shape of $\Omega_\sigma$ in the top-right panel of Fig.~\ref{fig:Background}. For the same values of $\{\sigma_i, \, V_0\}$, a larger $\xi$ sizeably increases the energy that the scalar field injects into the cosmic fluid once it starts to roll down its potential, an effect which, at a fixed value of $\xi$, can also be obtained by increasing the initial value of the scalar field $\sigma_i$. On the other hand, for larger values of $\xi$, we observe that $\Omega_\sigma$ becomes gradually more negative, therefore suppressing  $H(z)$, with respect to the $\xi=0$ case, before $\sigma$ starts to thaw, reducing the degeneracy  of the non-minimal coupling $\xi$ with the initial condition $\sigma_i$ (see also next Subsection).
Therefore our model offers a broader phenomenology than EDE ones, which is interesting since
the exact shape in redshift of the energy injection plays a crucial role in physical models that aim at solving the $H_0$ tension \cite{Knox:2019rjx}.
We stress that having $\Omega_\sigma<0$ is not a physical problem as $\Omega_\sigma$ only parmeterizes the contribution of the scalar field to the total expansion rate when the Friedmann equations are recast in \emph{Einstein gravity} form \cite{Boisseau:2000pr,Gannouji:2006jm}.  Although the main focus of this paper is the $\xi\geq0$ regime, it is also instructive to show the behavior of $\Omega_\sigma$ when the coupling is negative. We take the conformal coupling $\xi=-1/6$ as an example (see also Section \ref{sec:conf}). For such a large and negative $\xi$, the profile of the energy injection is continuous and resembles the one in models with extra dark radiation, exactly as the massless case with $\lambda=0$ \cite{Rossi:2019lgt,Ballesteros:2020sik,Braglia:2020iik}.

We stress again
that the quartic potential drastically modifies the scalar field evolution compared to the case in which $\lambda=0$. 
By the addition of the effective  mass, consistency of $G_{\rm eff}$ and PN parameters with Cavendish-type measurements and Solar System constraints, respectively, can be obtained without any fine tuning for $\xi >0$, as can be seen from the bottom panels of Fig.~\ref{fig:Background}. 
Note also that, thanks to the potential $V(\sigma)$, we have that $G_{\rm eff}$ grows with time, which is not possible in standard scalar-tensor models involving only the coupling $F(\sigma)$, for which $G_{\rm eff}$ decreases with time regardless of the sign of the non-minimal coupling \cite{Rossi:2019lgt,Braglia:2020iik}. However, in the conformally coupled case, $G_{\rm eff}$ decreases as in the massless case \cite{Braglia:2020iik}.

\subsection{Imprints of the Non-Minimal coupling on CMB and LSS}

\begin{figure}[h!]
		\begin{center} 
			\includegraphics[width=.49\columnwidth]{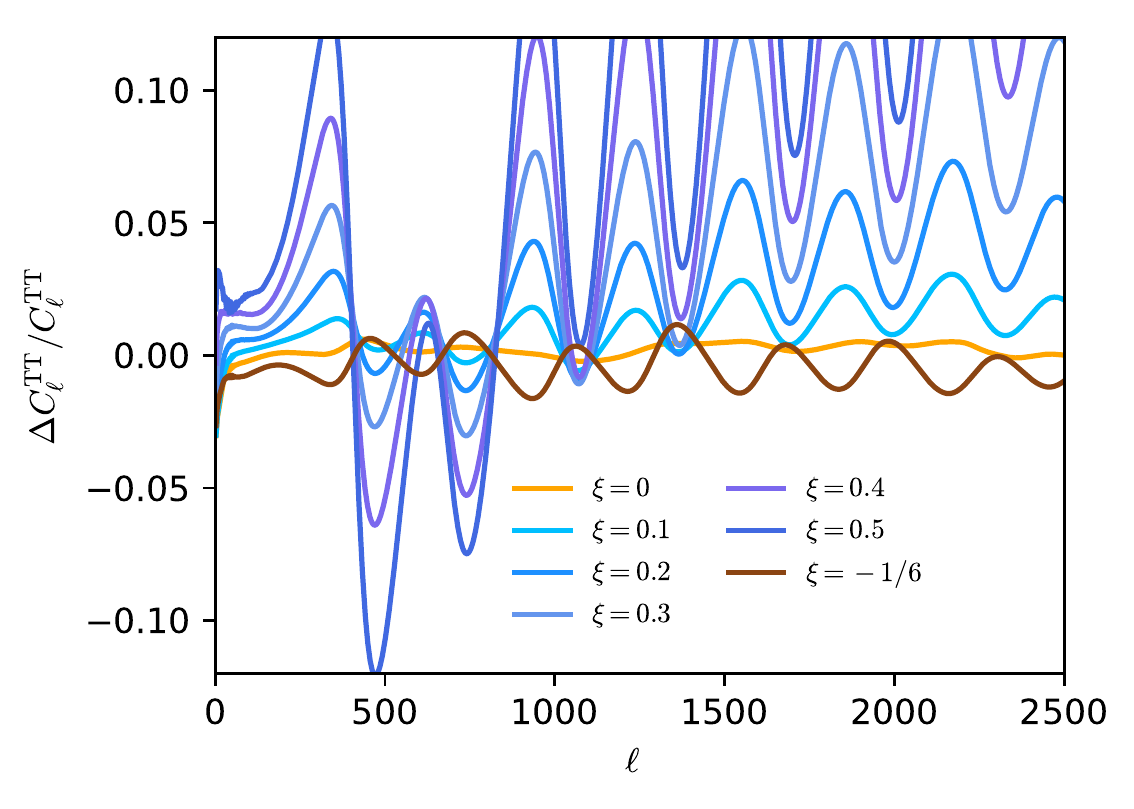}
			\includegraphics[width=.49\columnwidth]{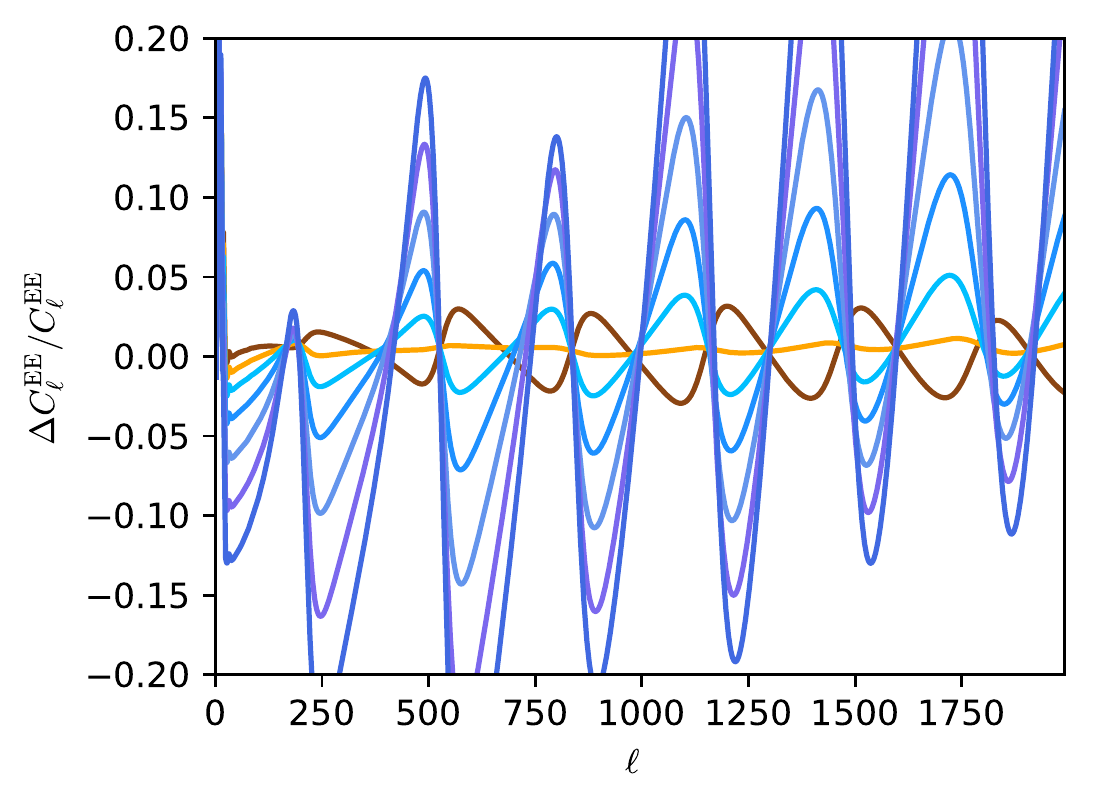}	
			\includegraphics[width=.49\columnwidth]{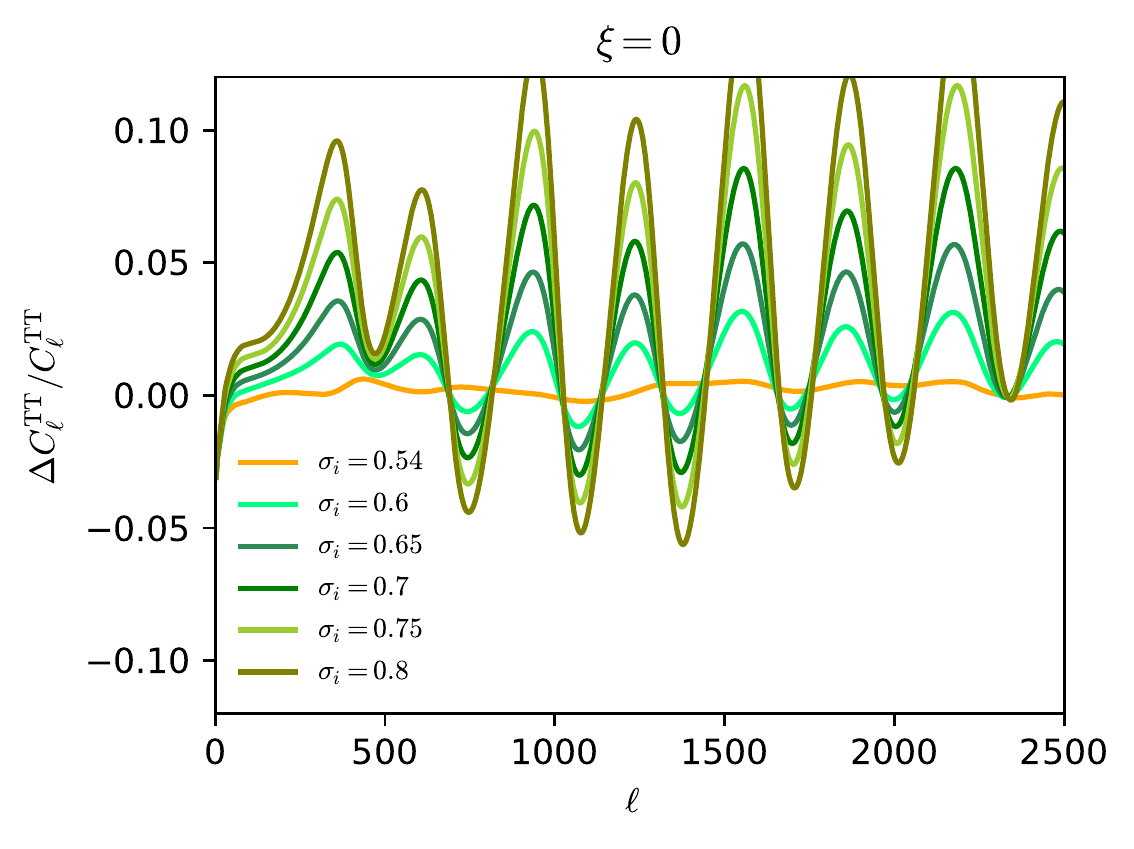}
			\includegraphics[width=.49\columnwidth]{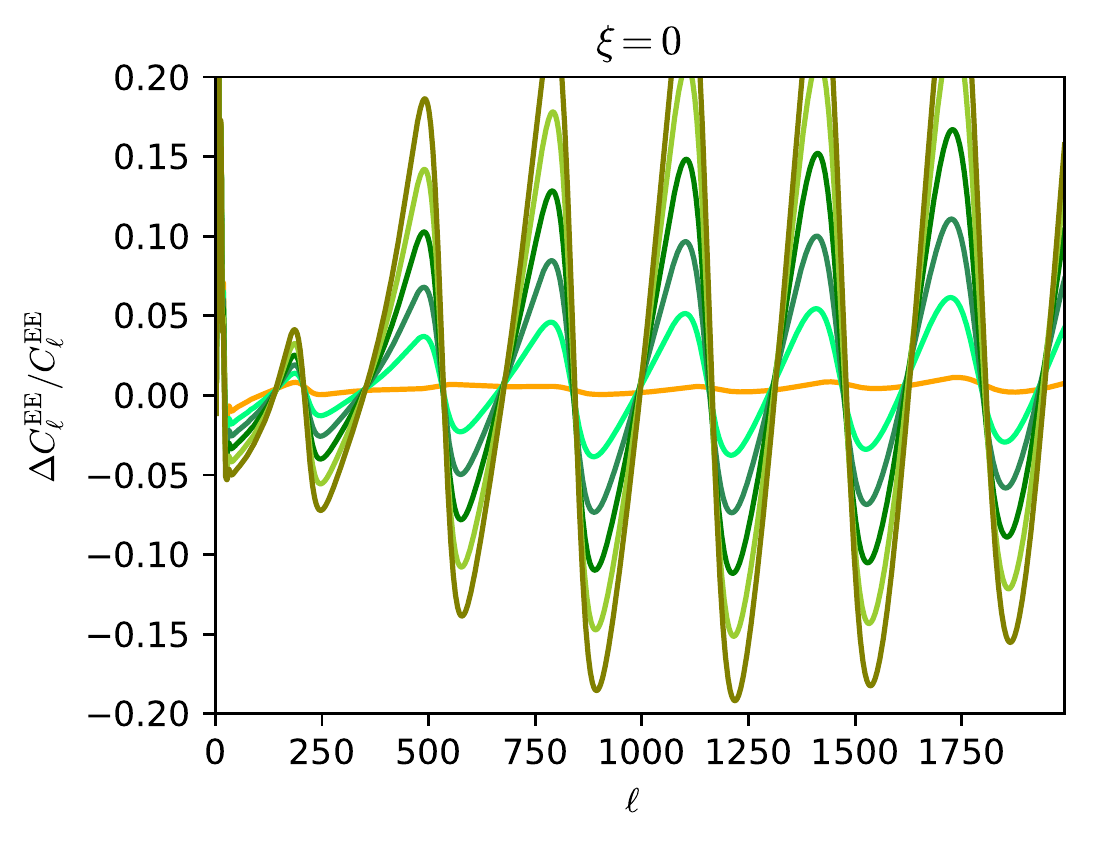}	
			\includegraphics[width=.49\columnwidth]{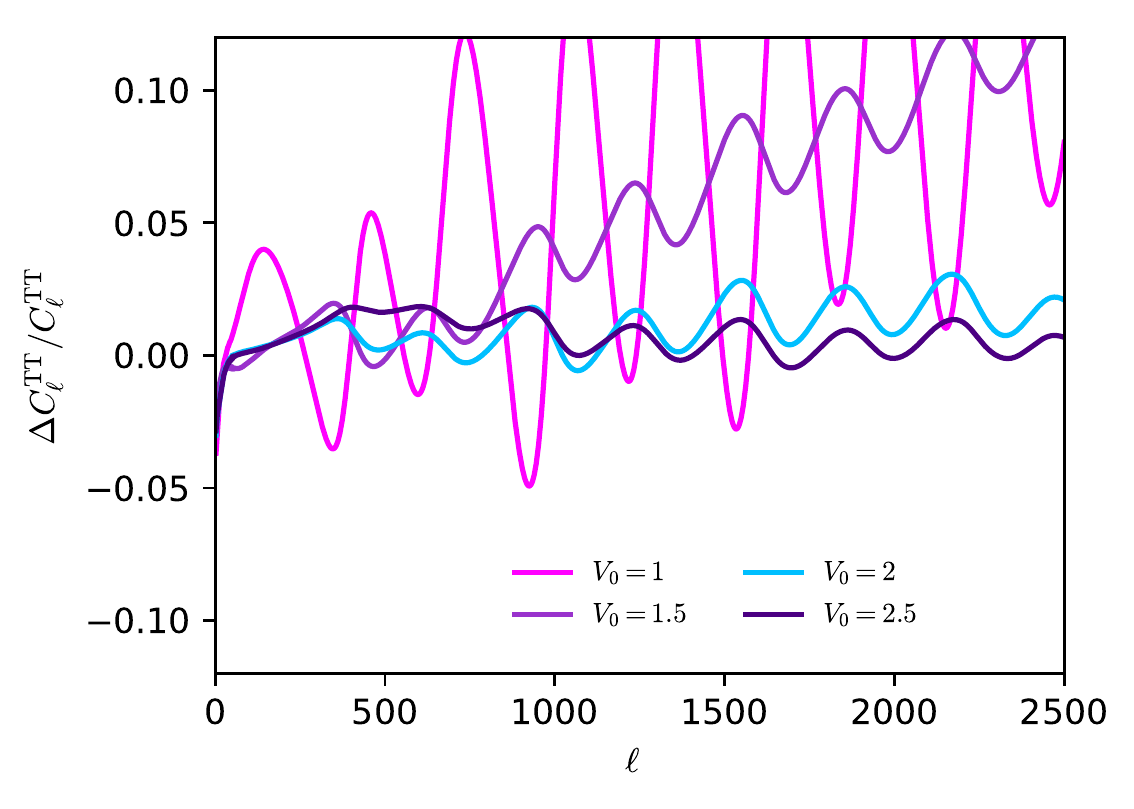}
			\includegraphics[width=.49\columnwidth]{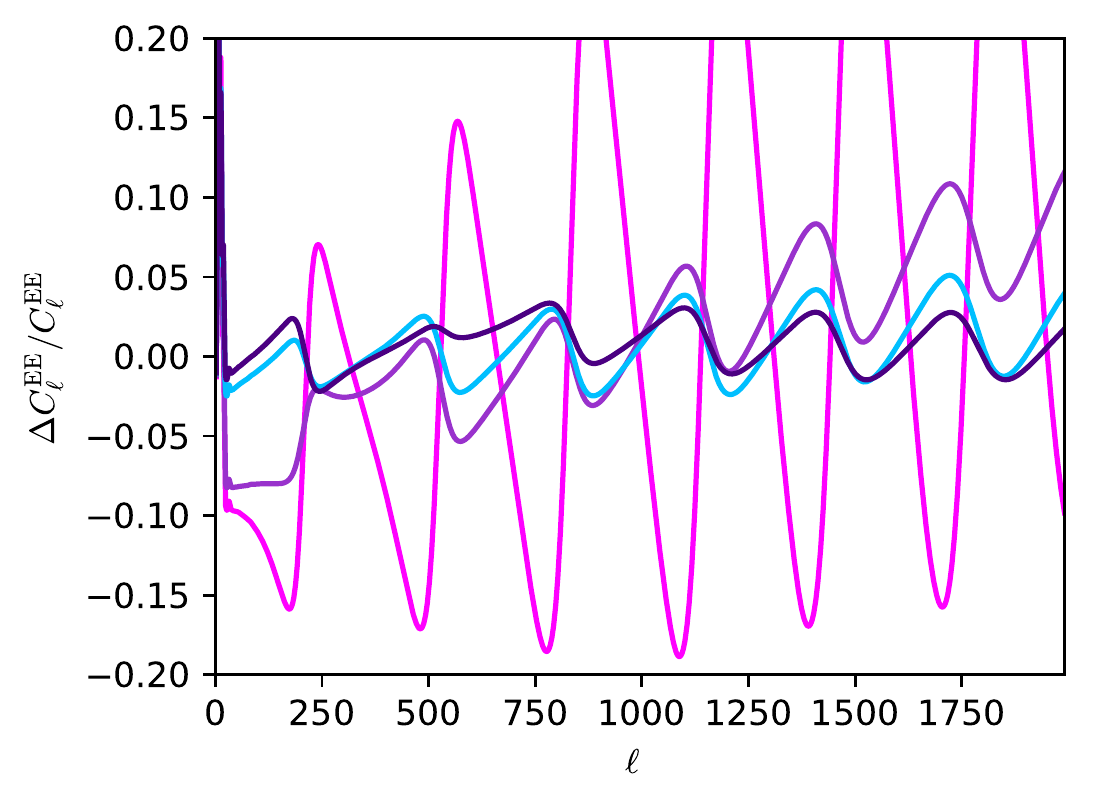}	
		\end{center}
		\caption{\label{fig:CMB} 
		 [Top] We plot the lensed CMB TT (left) and EE (right) angular power spectrum as a function of the non-minimal coupling $\xi$. [Center] We plot the lensed CMB TT (left) and EE (right) angular power spectrum as a function of the initial condition on the scalar field $\sigma_i$ with $\xi=0$. [Bottom] We plot the lensed CMB TT (left) and EE (right) angular power spectrum as a function of the potential parameter $V_0$ keeping the non-minimal coupling fixed to $\xi=0.1$.  We utilize the set of parameters used to produce Fig.~\ref{fig:Background}. }
	\end{figure}

We now show the imprints of EMG on CMB and LSS observables. 
The  temperature and E-mode polarization CMB angular power spectra are shown in the top panels of Fig.~\ref{fig:CMB}, from which it can be seen that the coupling sizeably affects the acoustic peaks structure of the CMB spectra, as a consequence of the modification to gravity around recombination. However, note that thanks to the potential $V(\sigma)$ and the different cosmological evolution of $\sigma$, the imprint of $\xi$ is drastically reduced with respect to the massless case with $\lambda=0$. Indeed, in the latter case, relative changes in $\Delta C_\ell/C_\ell$ of the same magnitude of the ones shown in the top panels of  Fig.~\ref{fig:CMB} can be obtained with much smaller values of $\xi$, see e.g. Fig.~9 of Ref.~\cite{Rossi:2019lgt}. We also note that the modifications to acoustic peaks 
for $\xi=-1/6$ are out of phase with respect to the $\xi>0$ case.

As discussed, the non-minimal coupling $\xi$ enhances the energy injection of the scalar field into the cosmic fluid, similarly to what can be obtained with a larger $\sigma_i$. In order to compare the two effects, in the bottom panels of Fig.~\ref{fig:CMB}, we fix $\xi=0$ and plot the residual CMB spectra  for a set of initial conditions $\sigma_i$ that give the same maximum energy injection of the curves presented in the top panel. Although both parameters modify the acoustic structure of the CMB, the pattern of the CMB residuals is different. In particular, given the same energy injection obtained by varying $\xi$ or $\sigma_i$ with $\xi=0$ respectively, the former has a stronger impact on the CMB since, thanks to the non-minimal coupling, the scalar field modifies the expansion history already while it is frozen, slightly decreasing $H(z)$ since its effective energy density is negative (see Fig.~\ref{fig:Background}).

\onecolumngrid

\twocolumngrid

			\begin{figure}
		\begin{center} 
			\includegraphics[width=\columnwidth]{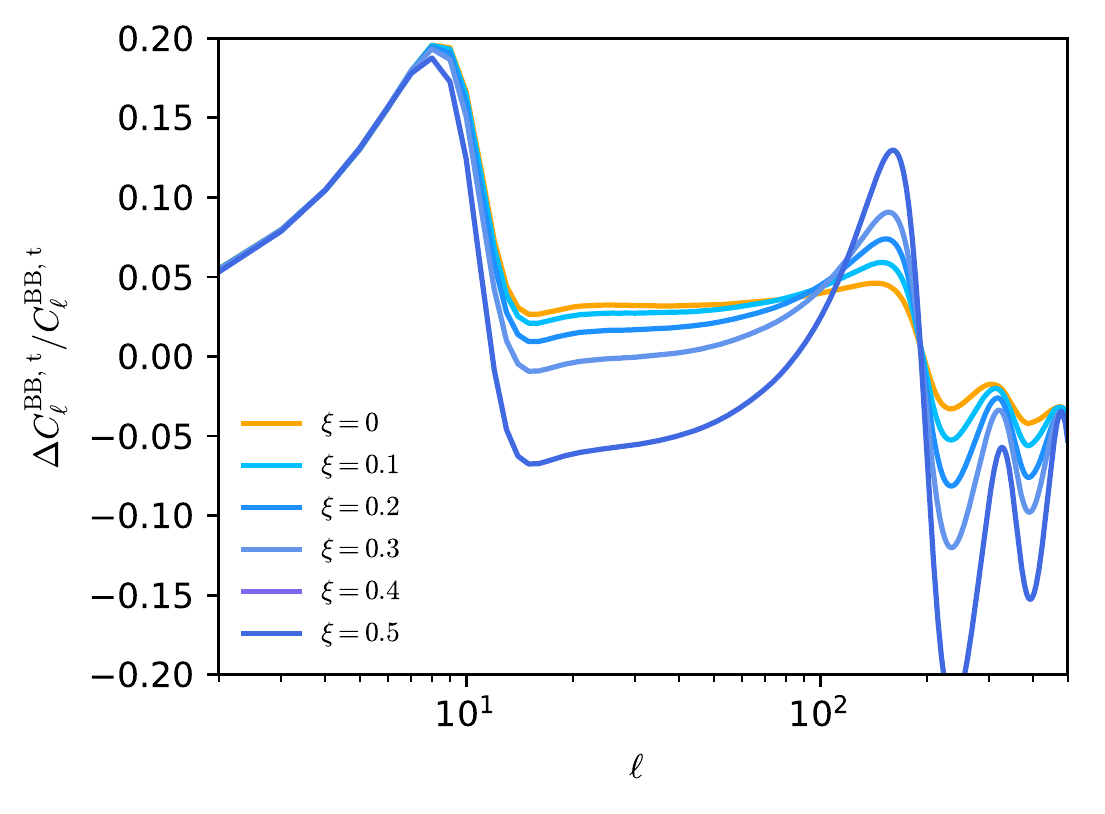}
			\includegraphics[width=\columnwidth]{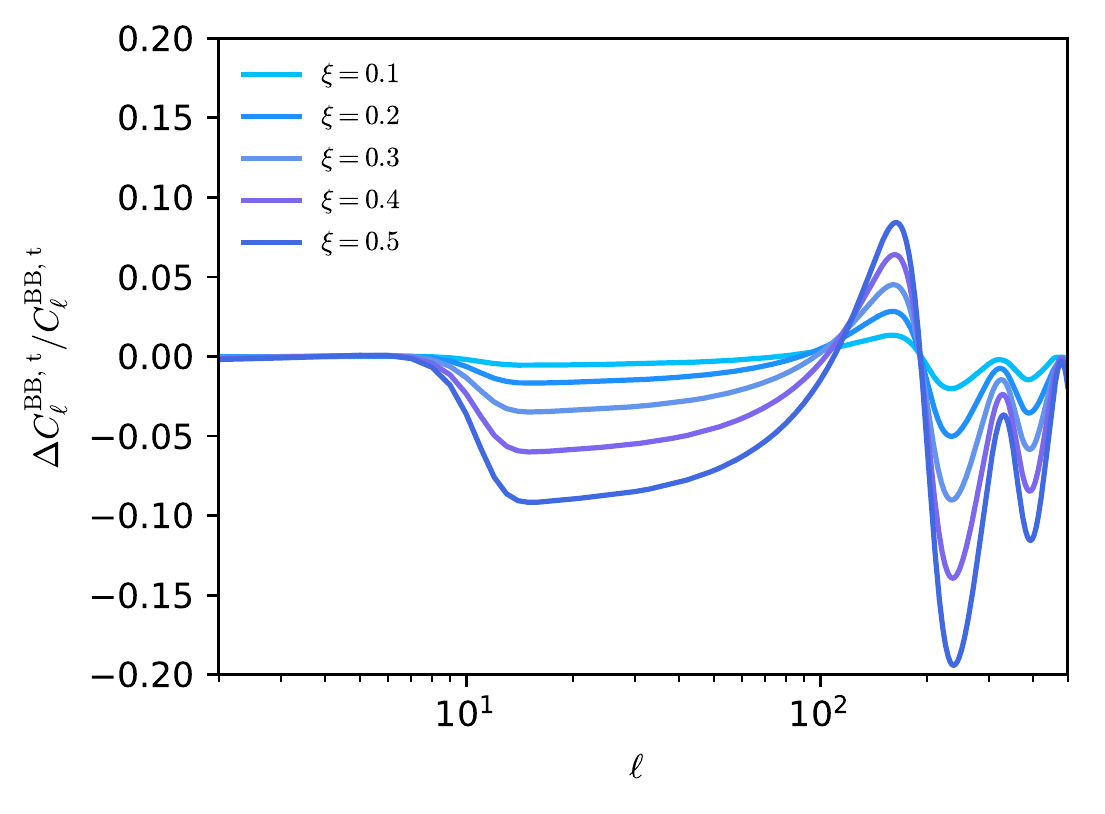}
		\end{center}
		\caption{\label{fig:Bmodes} 
		  We plot the CMB BB angular power spectrum due to tensor perturbations. In order to clarify the distinction between effects due to the shift in cosmological parameters and the genuine effects of the non-minimal coupling, we  plot both the relative differences between the EDE and $\Lambda$CDM baselines in Eq.~\eqref{eq:paramsRnR} and \eqref{eq:paramsLCDM} (top) and the ones obtained by fixing the EDE parameters in Eq.~\eqref{eq:paramsRnR} and varying $\xi$ (bottom). We set the tensor-to-scalar ratio to $r_{0.05}=0.05$.  
		  The lensing spectra are almost unaffected by varying our parameters, so the relative differences for the total spectra do not change from the ones in our plots.  }
	\end{figure}

In the perspective of future experiments dedicated to CMB polarization, it is also instructive to show the imprints of EMG on primordial B-mode polarization.
These are shown in Fig.~\ref{fig:Bmodes}, where we vary $\xi$ in the top panel. In the bottom panel, in order to better understand which effects are due to the shift of the cosmological parameters in Eqs.~\eqref{eq:paramsRnR} and \eqref{eq:paramsLCDM} and the ones of varying the coupling, we also plot the residuals obtained by fixing the cosmological parameters to Eq.~\eqref{eq:paramsRnR} and changing the coupling $\xi$   and fix $\xi=0$ and vary $\sigma_i$ in the right one. Note that a non-minimal coupling modifies the propagation equation for the two polarization states of the gravitational waves $h_{+,\,\times}$ as (neglecting anisotropic stresses for simplicity)\footnote{The term $\dot{F}/H F$ is sometimes referred to as $\alpha_M$  or $\delta$ in the literature \cite{Bellini:2014fua,Belgacem:2019pkk}.}:
\begin{equation}
    \ddot{h}_{+,\,\times}+\left(3 +\frac{\dot{F}}{H F}\right)H\dot{h}_{+,\,\times}+\frac{k^2}{a^2}=0.
\end{equation}
As shown in Refs.~\cite{Riazuelo:2000fc,Amendola:2014wma,Pettorino:2014bka}, the additional friction term induced by the non-minimal coupling may leave interesting observational signatures. In the case of $V(\sigma)=0$, the impact on B-mode polarization was analyzed in Ref.~\cite{Rossi:2019lgt}, where it was found the effects increase with $|\xi|$.
In our model, where the potential $V(\sigma)$ enlarge the range of $\xi$ which is compatible with the data (see next Section), the effects can indeed be larger, as can be seen from the left panel of Fig.~\ref{fig:Bmodes}. The effect of an increasing $\xi$ is twofold. First it changes the acoustic structure of the $C_\ell$'s for $\ell\gtrsim 100$, with a pattern which cannot be mimicked by a change in $\sigma_i$, similarly to what happens with the other CMB spectra, as can be appreciated by looking at the right panel of Fig.~\ref{fig:Bmodes}. Second, it also decreases the power in the range $10\lesssim \ell\lesssim 100$ compared to the $\Lambda$CDM model. Our plots also show a bump at very large scales. This, however, is a feature which is not directly ascribed to the EMG model or the EDE one. In fact, such a peak comes from the interplay of the different cosmological parameters in Eqs.~\eqref{eq:paramsRnR} and \eqref{eq:paramsLCDM}. Nevertheless, we have verified that such a bump also occurs when considering the relative differences between the bestfit values for $\Lambda$CDM and EMG/EDE cosmologies shown in the next Section, and thus it may constitute an indirect  signature of EMG and EDE models that can be tested with future CMB B modes experiments.

		\begin{figure}
		\begin{center} 
			\includegraphics[width=\columnwidth]{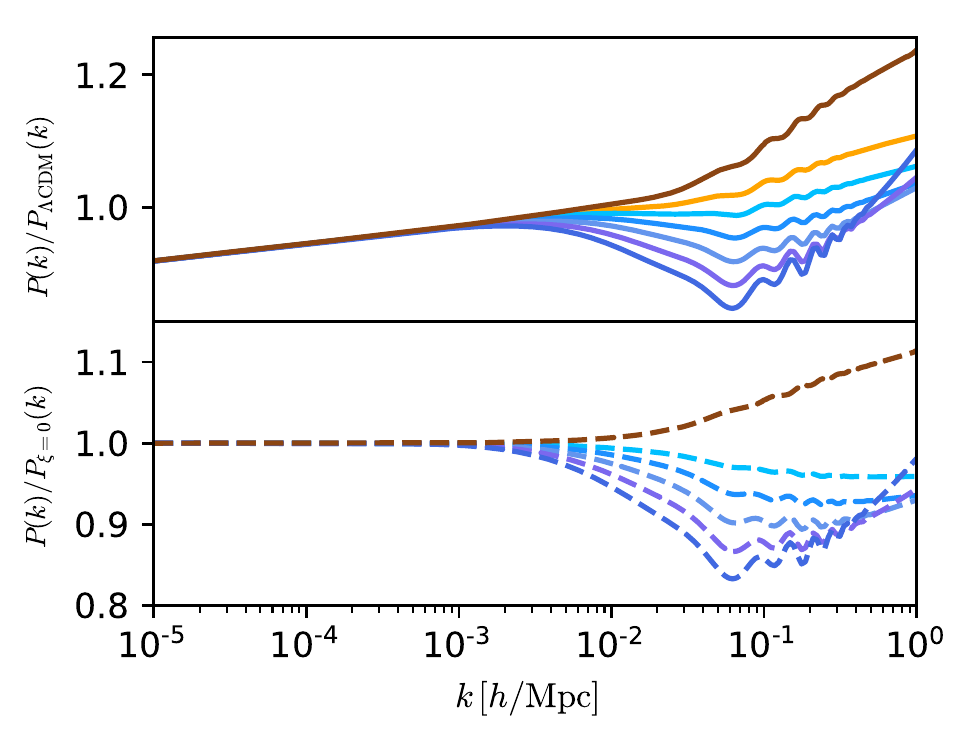}
			\includegraphics[width=\columnwidth]{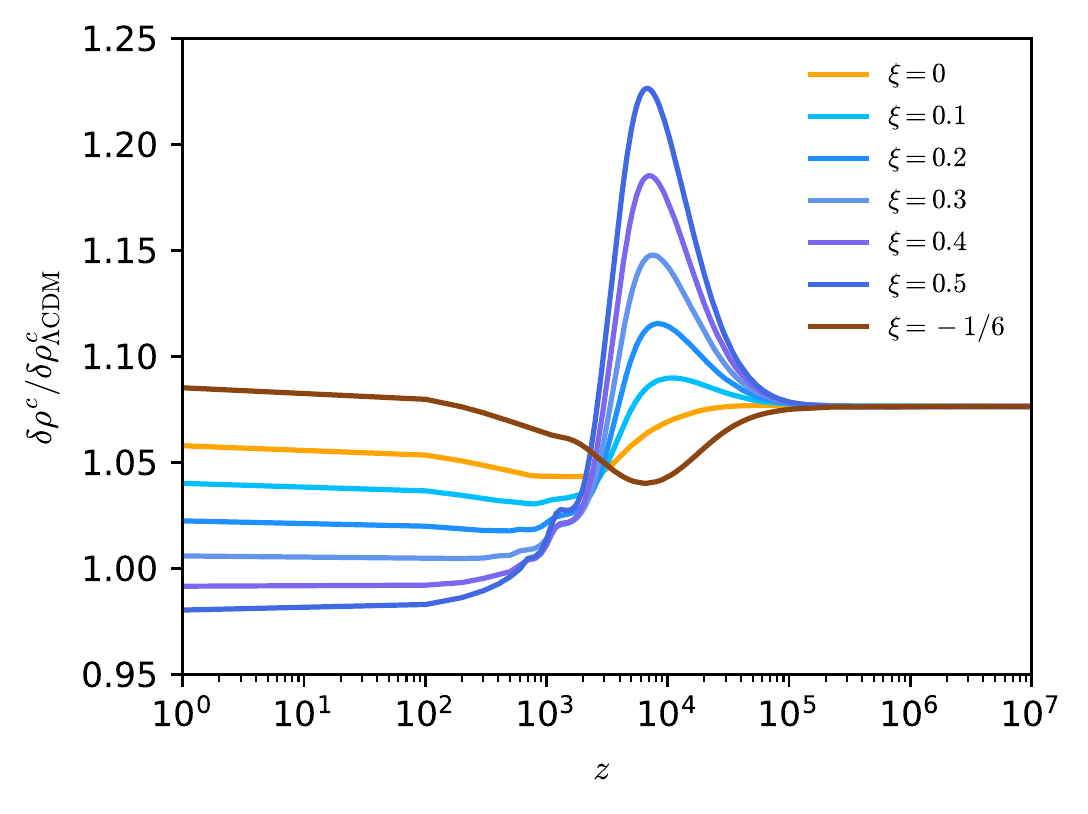}	
		\end{center}
		\caption{\label{fig:Pk} 
		 We plot  the ratio of the EMG and $\Lambda$CDM linear matter power spectra at $z=0$ (top) and the evolution of the dark matter perturbation $\delta\rho_c$ for $k=0.1$ $h$/Mpc divided by the one for the $\Lambda$CDM model (bottom) as a function of the non-minimal coupling $\xi$. As in the previous plot, for solid lines, we utilize the set of parameters used to produce Fig.~\ref{fig:Background} and we compute relative differences between the EDE and $\Lambda$CDM baselines in Eq.~\eqref{eq:paramsRnR} and \eqref{eq:paramsLCDM}. In the lower half of the top panel, in order to make clear which are the effects due only to the variation of $\xi$, we  fix  the EDE baseline parameters in Eq.~\eqref{eq:paramsRnR} and plot $P(k)$  varying $\xi$  using dashed lines. }
	\end{figure}
	
Since EDE scenarios have been recently shown to be constrained by the matter power spectrum at low redshift \cite{Hill:2020osr,Ivanov:2020ril,DAmico:2020ods,Bull:2020cpe,Klypin:2020tud}, it is important to investigate the imprints of our model also on LSS and compare them to the ones of NMC and EDE models. We plot the ratio between the linear matter power spectra for our EMG model and the $\Lambda$CDM one in the left panel of Fig.~\ref{fig:Pk}. 
As previously studied in \cite{Umilta:2015cta,Ballardini:2016cvy,Nunes:2018xbm,Lin:2018nxe,Rossi:2019lgt}, the matter power spectrum is enhanced at small scales in effectively massless scalar-tensor models aiming at alleviating the $H_0$ problem since gravity was relatively stronger at early times.
Analogously, EDE models also enhance the matter power spectrum at small scales compared to the $\Lambda$CDM one, as can be seen from the orange line in the plot. It is however important to understand that this effect is not due to the EDE component itself, but rather by the shift towards a larger $\omega_c$ that is needed to maintain the fit to the CMB data, see Eqs.~\eqref{eq:paramsRnR} and \eqref{eq:paramsLCDM}. In fact, the larger is the fraction of EDE, the greater is the suppression of the growth of the perturbations within the horizon during the epoch when EDE is not negligible. We see from the top panel of Fig.~\ref{fig:Pk} that, fixing all the other parameters, the non-minimal coupling $\xi$ goes instead in the direction of suppressing the power at small scales, as it weakens the strength of gravity  during the EMG epoch, see Fig.~\ref{fig:Background}. This is not true anymore for the $\xi=-1/6$ case in which a stronger gravity ($G_{\rm eff}/G>1$) at early times leads to an enhancement of the power at smaller scales. Again, the results are completely different from the case with $\lambda=0$, for which the $G_{\rm eff}$ always decreases with time, leading to a stronger gravity at early times and a consequent larger power in $P(k)$ at small scales \cite{Rossi:2019lgt}.  

The results in Fig.~\ref{fig:Pk} can be better understood by looking at the evolution of dark matter perturbations. For this purpose, we plot the evolution of the ratio of the dark matter perturbation $\delta\rho_c$ for the EMG and the $\Lambda$CDM model for the mode $k=0.1$ $h/$Mpc in the bottom panel of Fig.~\ref{fig:Pk}.  As can be seen, for a positive $\xi$, initially scalar field perturbations enhance the growth of dark matter perturbations with respect to the $\Lambda$CDM case, overcoming the suppression factor due to having $G_{\rm eff}/G<1$. The opposite occurs for a negative value, as can be seen from the brown line. On even smaller scales (larger $k$), we also have a fifth force (scale dependent) contribution from the scalar field perturbations that further enhances the growth of dark matter perturbations at very early times with respect to the $\Lambda$CDM case, which explains the raise in the $ P(k)$ at 
small scales for $\xi=0.5$ in the left panel of Fig.~\ref{fig:Pk}.

Once the scalar field starts to roll down the potential, however, the scalar field perturbations become negligible and the only effect of the modification to gravity is to suppress (enhance) the gravitational potentials by a factor of $F(\sigma)<1$ ($>1$) depending on the sign of $\xi$, leading to the observed suppression (enhancement) in the left panel of Fig.~\ref{fig:Pk}.

		\begin{figure}
		\begin{center} 
			\includegraphics[width=\columnwidth]{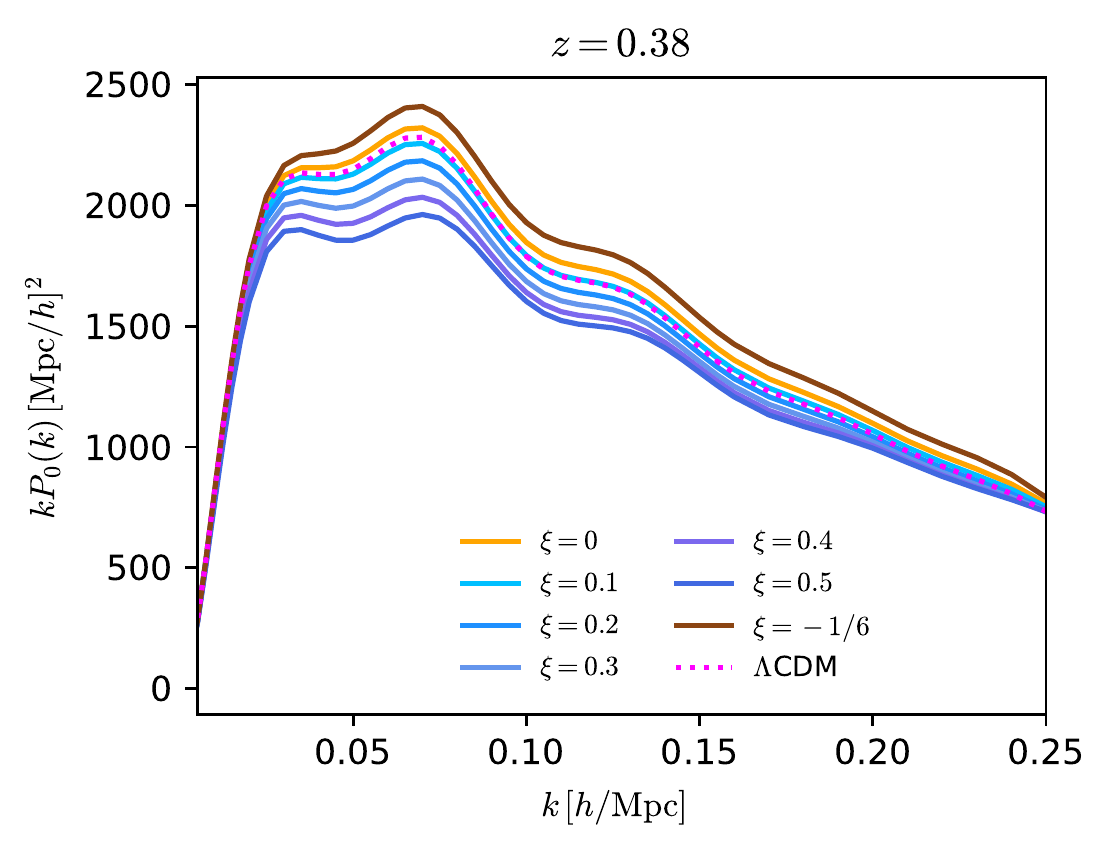}
			\includegraphics[width=\columnwidth]{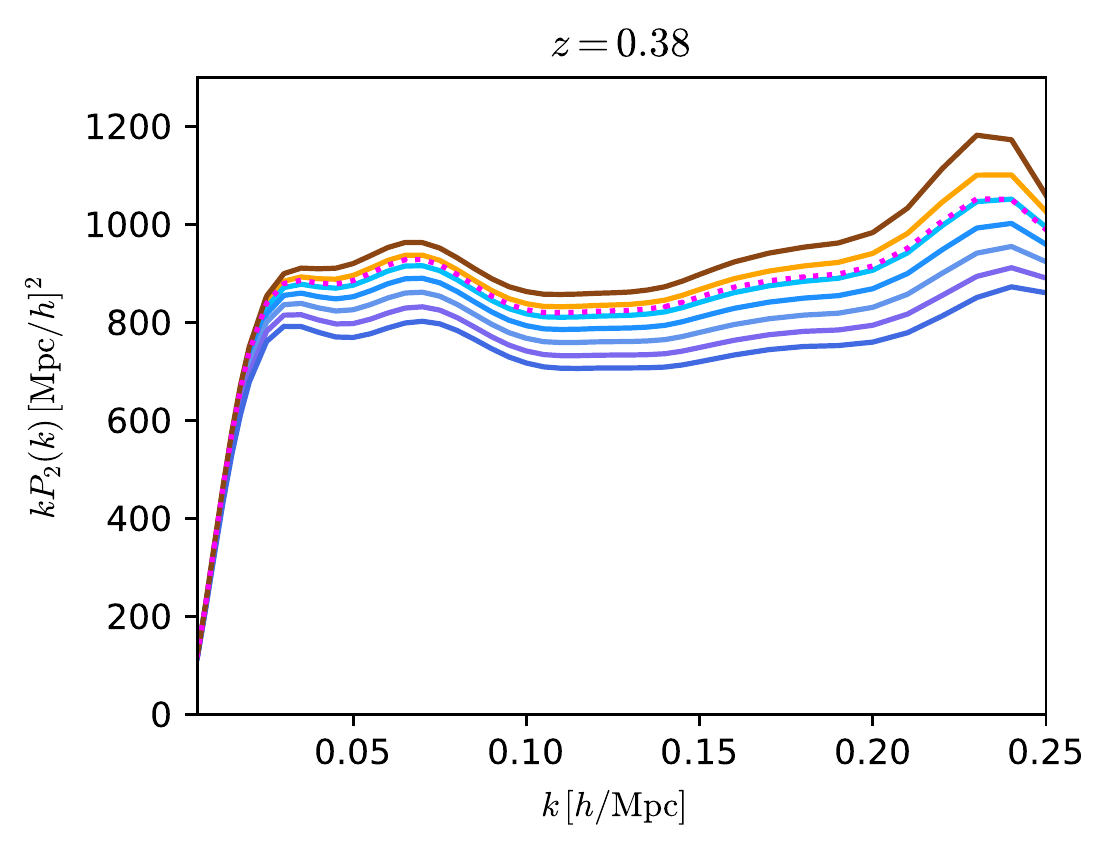}
		\end{center}
		\caption{\label{fig:Plk} 
 We plot the 1 loop $\ell=0$ (left) and $\ell=2$ (right) multipole moments of the galaxy power spectrum  as a function of the non-minimal coupling $\xi$. We utilize the set of parameters used to produce Fig.~\ref{fig:Background}. We also plot the $\Lambda$CDM results in magenta dotted lines for a comparison. }
	\end{figure}

Furthermore, it is instructive to show the effects on the observed redshift-space galaxy-spectrum. We plot the multipole moments in  Fig.~\ref{fig:Plk}  where we show the monopole $\ell=0$ (left panel) and the quadrupole $\ell=2$  (right panel) resummed at 1 loop order in perturbation theory, which we have produced with the publicly available code PyBird\footnote{\href{https://github.com/pierrexyz/pybird}{https://github.com/pierrexyz/pybird}} \cite{DAmico:2020kxu}. 
Although PyBird works in the framework of a $\Lambda$CDM effective field theory of LSS, 
the deviations from General Relativity at the relevant redshift considered by PyBird are so small that its use in this context is safe (see Fig.~\ref{fig:Background}).  As an example we have considered the multipole moments at $z=0.38$, which corresponds to the redshift of the low-$z$ NGC BOSS data (see next Section). Note that the effect of $\xi$ is to reduce the amplitude of both $P_0(k)$ and $P_2(k)$. It is very interesting to note that, starting from the parameters in Eqs.~\eqref{eq:paramsRnR} and \eqref{eq:paramsLCDM}, we recover very similar spectra for $\Lambda$CDM and the EMG model with $\xi=0.1$, suggesting that the non-minimal coupling can help reconcile EDE models with LSS observations. 

\section{Methodology and datasets}
\label{sec:data}
We perform a Markov-chain Monte Carlo (MCMC) analysis using a modified version of the hiCLASS code \cite{Lesgourgues:2011re,Blas:2011rf,Zumalacarregui:2016pph,Bellini:2019syt} interfaced to the publicly available sampling code MontePython-v3\footnote{\href{https://github.com/brinckmann/montepython\_public}{https://github.com/brinckmann/montepython\_public}} 
\cite{Audren:2012wb,Brinckmann:2018cvx} and to the PyBird code for the calculation of the full shape of the galaxy power spectrum in the effective field theory of large scale structure \cite{DAmico:2020kxu}. We obtain plots using the GetDist package\footnote{\href{https://getdist.readthedocs.io/}{https://getdist.readthedocs.io/en/latest}} \cite{Lewis:2019xzd}. For all our runs, we set the initial velocity of the scalar field to zero, use adiabatic initial conditions for the scalar field perturbations \cite{Rossi:2019lgt,Paoletti:2018xet} and consider massless neutrinos ($N_{\rm eff}=3.046$)\footnote{We have tested that the differences with respect to having one massive neutrino with $m_\nu=0.06$ eV  in the estimate of the cosmological parameters are small except for a shift in $H_0$ towards larger values and a smaller $\sigma_8$  (though $S_8$ does not change appreciably).}

We sample over the cosmological parameters $\{\omega_b,\,\omega_{cdm}, \,\theta_s,\,\ln 10^{10}A_s,\,n_s,\,\tau_\textup{reio},\,\xi,\,\sigma_i,\,V_0\}$ using the Metropolis-Hastings algorithm and with a Gelman-Rubin \cite{Gelman:1992zz}  convergence criterion $R-1<0.03$. For the extra parameters we consider flat priors $\xi\in [0,1]$, $\sigma_i/M_{\rm pl}\in[0, 0.9]$ and $V_0\in[0.6,\,3.5]$. Note that EDE models are usually parametrized with two parameters describing the redshift at which the scalar field starts to roll down the potential, usually denoted as critical redshift $z_c$, and the maximum energy injection $f_{\rm scf}$ \cite{Poulin:2018dzj,Poulin:2018cxd,Agrawal:2019lmo}. For the particular case of the RnR model, the correspondence between $\{V_0,\,\sigma_i\}$ and $\{z_c,\,f_{\rm scf}\}$ is unique under the assumption of the same initial velocity of the scalar field. However, as explained in Section~\ref{sec:model}, this one to one correspondence is not possible in our model, where also $\xi$ contributes to the energy injection into the cosmic fluid.  For this reason, 
we prefer to use the physical parameters describing our model \eqref{eq:model} as previously done \cite{Braglia:2020iik}. Nevertheless, we quote $\log_{10}\,z_c$ and $f_{\rm scf}\equiv\Omega_{\rm scf}$ as derived parameters. 
Note that we model the non-linear power spectra using HALOFIT \cite{Smith:2002dz,Takahashi:2012em}. In this respect, see also Ref.~\cite{Murgia:2020ryi} for a comparison between of HALOFIT and HMcode~\cite{Mead:2015yca} in the context of EDE.

For each run, we also compute the best-fit values extracted  using the MINUIT algorithm \cite{James:1975dr} implemented in the IMINUIT python package\footnote{\href{https://iminuit.readthedocs.io/}{https://iminuit.readthedocs.io/en/stable/}} and quote the difference in the model $\chi^2$ with respect to $\Lambda$CDM one, i.e. $\Delta \chi^2 = \chi^2 - \chi^2 (\Lambda\mathrm{CDM})$, where negative values indicate an improvement in the fit of the given model with respect to the $\Lambda$CDM for the same dataset.

In order to quantify to what extent the improvement in the fit to the data warrants the increase in the model complexity compared to the baseline $\Lambda$CDM model, we compute the Bayes factor defined as the ratio of the evidences for the extended model ${\cal M}_{\rm E}$ with respect to the baseline ${\cal M}_{\rm L}$ as \cite{Jeffreys:1939xee}:

\begin{eqnarray}
B_{EL} \equiv \frac{\int d\boldsymbol{\theta}_E\, \pi(\boldsymbol{\theta}_E \vert {\cal M}_E) {\cal L}(\mathbf{x} \vert \boldsymbol{\theta}_E,{\cal M}_E)\,,}{\int d\boldsymbol{\theta}_L\, \pi(\boldsymbol{\theta}_L \vert {\cal M}_L) {\cal L}(\mathbf{x} \vert \boldsymbol{\theta}_L,{\cal M}_L)\,,}\,,
\label{eq:bayesfactor}
\end{eqnarray}
where $\pi(\boldsymbol{\theta}_{\rm E,\,L})$ is the prior for the parameters ${\theta}_{\rm E,\,L}$ and ${\cal L}(\mathbf{x} \vert \boldsymbol{\theta}_{\rm E,\,L})$ the likelihood of the data given the model ${\cal M}_{\rm E,\,L}$. The extent to what the extended model ${\cal M}_{\rm E}$ is preferred over the baseline ${\cal M}_{\rm L}$ can be qualitatively assessed using the Jeffreys scale reported in Table~\ref{tab:kassraftery} \cite{Kass:1995loi}.
We compute the evidence directly from our MCMC using the method introduced in Ref.~\cite{Heavens:2017afc} implemented in the MCEvidence code\footnote{\href{https://github.com/yabebalFantaye/MCEvidence}{https://github.com/yabebalFantaye/MCEvidence}}.

\begingroup
\begin{center}
\begin{table}[!h]
\begin{tabular}{|c||c|}
\hline
$\ln B\equiv\ln B_{\rm E L}$ & Strength of preference for model ${\cal M}_i$ \\ \hline \hline
$0 \leq \ln B < 1$ & Weak \\ \hline
$1 \leq \ln B < 3$ & Definite \\ \hline
$3 \leq \ln B < 5$ & Strong \\ \hline
$\ln B \geq 5$ & Very strong \\
\hline
\end{tabular}
\caption{Revised Jeffreys scale used to interpret the values of $\ln B$ obtained when comparing two competing models through their Bayesian evidence \cite{Kass:1995loi}. A value of $\ln B>0$ indicates that the extended model is favoured with respect to the $\Lambda$CDM baseline model. }
\label{tab:kassraftery}
\end{table}
\end{center}
\endgroup

We constrain the cosmological parameters using several combination of datasets. 
Our CMB measurements are those from the {\em Planck} 2018 legacy release (hereafter P18) 
on temperature, polarization, and weak lensing CMB anisotropies angular power spectra 
\cite{Aghanim:2019ame,Aghanim:2018oex}. 
The high-multipoles likelihood $\ell \geq 30$ is based on {\tt Plik} likelihood. 
We use the low-$\ell$ likelihood combination at $2 \leq \ell < 30$: temperature-only 
{\tt Commander} likelihood plus the {\tt SimAll} EE-only likelihood.
For the {\em Planck} CMB lensing likelihood, we consider the {\em conservative} 
multipoles range, i.e. $8 \leq \ell \leq 400$. We marginalize over foreground and 
calibration nuisance parameters of the {\em Planck} likelihoods \cite{Aghanim:2019ame,Aghanim:2018oex} 
which are also varied together with the cosmological ones. We refer to this CMB dataset as P18.

We use the baryon acoustic oscillation (BAO) of
Baryon Spectroscopic Survey (BOSS) DR12 \cite{Alam:2016hwk} post-reconstructed power spectrum measurements in three redshift slices with effective redshifts $z_{\rm eff} = 0.38,\,0.51,\,0.61$ 
\cite{Ross:2016gvb,Vargas-Magana:2016imr,Beutler:2016ixs}, 
in combination with the 'small-$z$' measurements from 6dF \cite{Beutler:2011hx} at $z_{\rm eff} = 0.106$ 
and the one from SDSS DR7 \cite{Ross:2014qpa} at $z_{\rm eff} = 0.15$. We refer to this combination of BAO data as BAO.

We also use the full shape of the BOSS DR12 pre-reconstructed galaxy clustering measurements \cite{Gil-Marin:2015sqa} using the Effective Field Theory (EFT) of LSS analysis of Refs.~\cite{DAmico:2019fhj,Colas:2019ret}. In particular we consider the combination of the monopole and quadrupole of the power spectra of the three different sky-cuts CMASS NGC and CMASS SGC at effective redshift $z_{\rm eff}=0.57$ and LOWZ NGC at $z_{\rm eff}=0.32$ and we follow the conventions of Refs.~\cite{DAmico:2019fhj,Colas:2019ret,DAmico:2020kxu} for the maximum wavenumber that we consider ($k_{\rm max}=0.23\,h/\text{Mpc}$ for CMASS and $k_{\rm max}=0.20\,h/\text{Mpc}$ for NGC). We combine this dataset with the $H r_s$ and $D_A/r_s$ parameters measured from the post-reconstructed power spectra corresponding to the same sky-cuts, see Ref.~\cite{DAmico:2020kxu} for an explanation of how the covariances between these datasets are calculated. We refer to such a dataset, combined with 'small-$z$' BAO mentioned in the previous paragraph as BAO+FS.

Additionally, we use the Pantheon supernovae dataset \cite{Scolnic:2017caz}, which includes measurements of the luminosity distances of 1048 SNe Ia in the redshift range $0.01< z <2.3$. We refer to Pantheon data as SN. As discussed in \cite{Braglia:2020iik}, we do not consider any corrections due to the change in the peak luminosity of SNe induced by the time evolution of the gravitational constant \cite{GarciaBerro:1999bq,Riazuelo:2001mg, Nesseris:2006jc, Wright:2017rsu}, since they should lead to a negligible effect for the models considered here.

We also consider the combination with a Gaussian likelihood based on the latest
determination of the Hubble constant from Hubble Space Telescope (HST) observations 
(hereafter R19), i.e. $H_0 = (73.5 \pm 1.4)$ km s$^{-1}$Mpc$^{-1}$ \cite{Reid:2019tiq} and from time delay from gravitationally lensed quasars
 from the H0LiCOW collaboration \cite{Wong:2019kwg}, that is 
$H_0 = \left(73.3_{-1.8}^{+1.7}\right)$ km s$^{-1}$Mpc$^{-1}$. Since there is no correlation between the two measurements, they can be combined again in an inverse-variance weighted Gaussian prior as $H_0 = \left(73.4\pm1.1\right)$ km s$^{-1}$Mpc$^{-1}$. We refer to this prior simply as $H_0$.

Finally, in order to include weak lensing data from photometric surveys, we follow Refs.~\cite{Hill:2020osr,Ivanov:2020ril} and implement them through a Gaussian prior on the parameter $S_8=\sigma_8\sqrt{\Omega_m/0.3}$. Compressing a large amount of data in a single data point obtained for $\Lambda$CDM is just an approximation and the correct way would be to perform a full fledged analysis with the correct likelihood. However, it was demonstrated in Ref.~\cite{Hill:2020osr}, that, when combined with CMB, BAO, SN, the DES-Y1 likelihood from two-point correlations of photometric galaxy clustering, galaxy-galaxy lensing and cosmic shear is well approximated by a Gaussian prior on $S_8=0.773\pm{0.026}$ in $\Lambda$CDM and EDE models. 
Note also that another thorough analysis of EDE models with weak lensing surveys is performed in Ref.~\cite{Murgia:2020ryi}. In that paper, the analysis with the full KiDS+VIKING likelihood and the one where the joint KiDS+VIKING and DES data are approximated by a Gaussian prior again show a qualitative agreement between the two approaches, although the joint KiDS+VIKING and DES data are somewhat more constraining.
Given these results, we use as a proxy for complementary measurements on $S_8$ from galaxy weak lensing a Gaussian prior for
the inverse-variance weighted combination of the measurements of DES \cite{Abbott:2017wau}, KV-450 \cite{Hildebrandt:2016iqg,Hildebrandt:2018yau} and HSC \cite{Hikage:2018qbn}, i.e. $S_8=0.770\pm0.017$.  We refer to this prior simply as $S_8$. We leave the task of a full weak lensing analysis to assess the reliability of our approximation to a future work. 

As a final comment, note that also Big Bang Nucleosynthesis data constrain the variation of the $G_N$ from the early times to today. As discussed more in depth in Refs.~\cite{Ballesteros:2020sik,Braglia:2020iik}, in our model this translates into a constraint on the quantity $\xi\sigma_i^2$, which is constrained to be   $\Delta G_N/G_N= (G_N(t_{\rm BBN})-G_N(t_0))/G_N(t_0)\simeq\xi\sigma_i^2=1.01^{0.20}_{-0.16}$ at a 68\% CL level \cite{Copi:2003xd,Bambi:2005fi}. A tighter constraint was also derived more recently in Ref.~\cite{Alvey:2019ctk}, which found $\Delta G_N=0.02\pm0.06$.  As we will show in the next Section, contraints from the data set introduced above are always tighter or nevertheless consistent with BBN ones, so we do not need to include BBN data in our analysis\footnote{In our model,  the $\sigma$ field is frozen deep in the radiation era. Although we do not consider this possibility in our paper, quantum mechanically, it is possible that the scalar field random walks and ends up at a larger values at earlier times. This leads to a larger $H$, which becomes in tension with BBN constraints.  A way out to this problem can be to assume a different  non-minimal coupling of the form $F(\sigma)=M_{\rm pl}^2+\xi\sin^2\left[f(\sigma)\right]$ so that it never gets too large.  }.

\section{Results} \label{sec:results}

In this Section, we present the results of our MCMC analysis performed using several combinations of the data sets introduced in the previous section. We comment on each combination in turn and  we only present plots of the posterior distribution of the cosmological parameters varied in the analysis and refer the reader to Appendix \ref{sec:appendix} for the tables containing their mean and best-fit values, as well as the $\chi^2$ for each data set and the Bayes factors.  

\begin{figure}
		\begin{center}
			\includegraphics[width=.7\columnwidth]{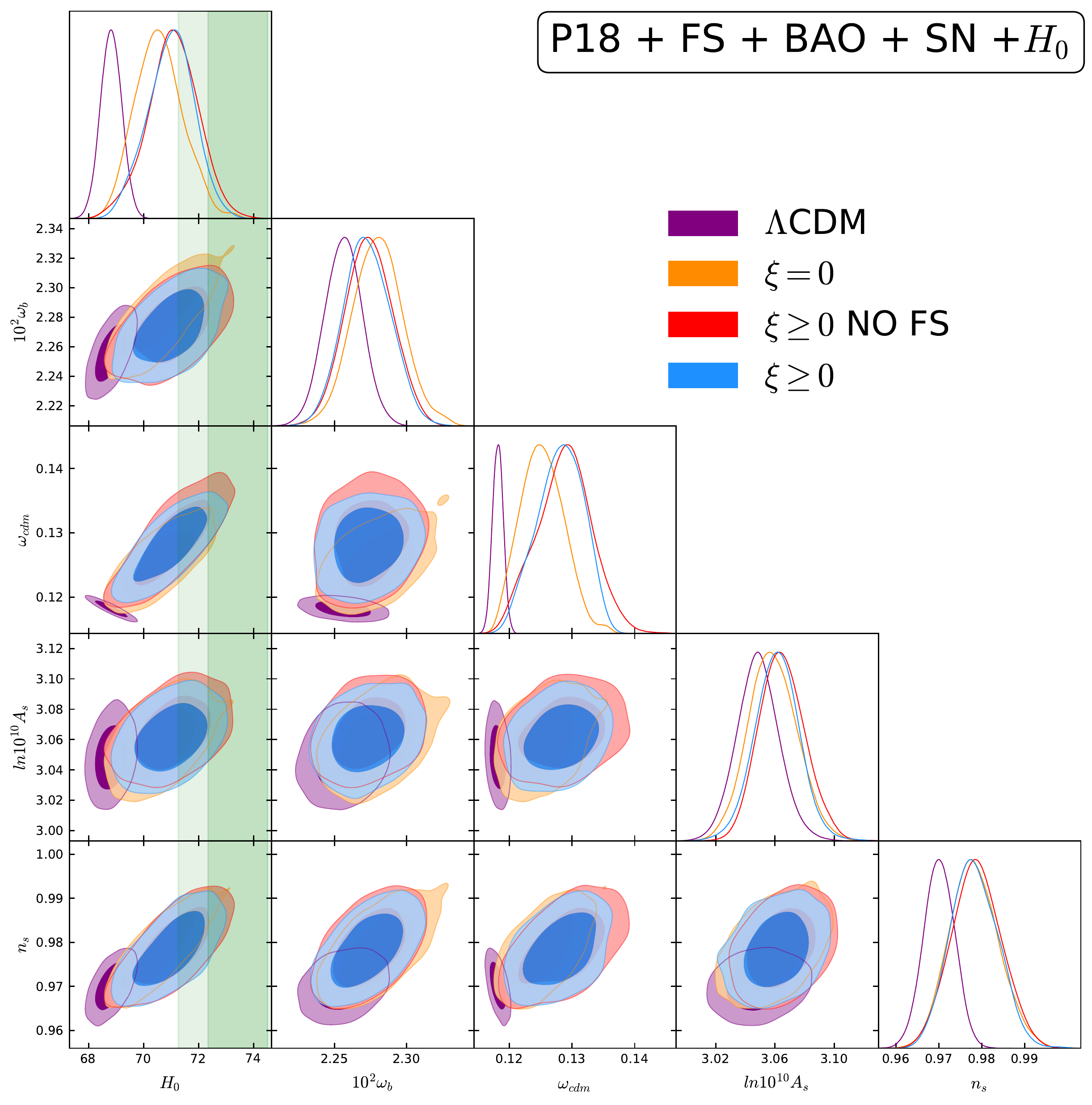}	
			\includegraphics[width=.29\columnwidth]{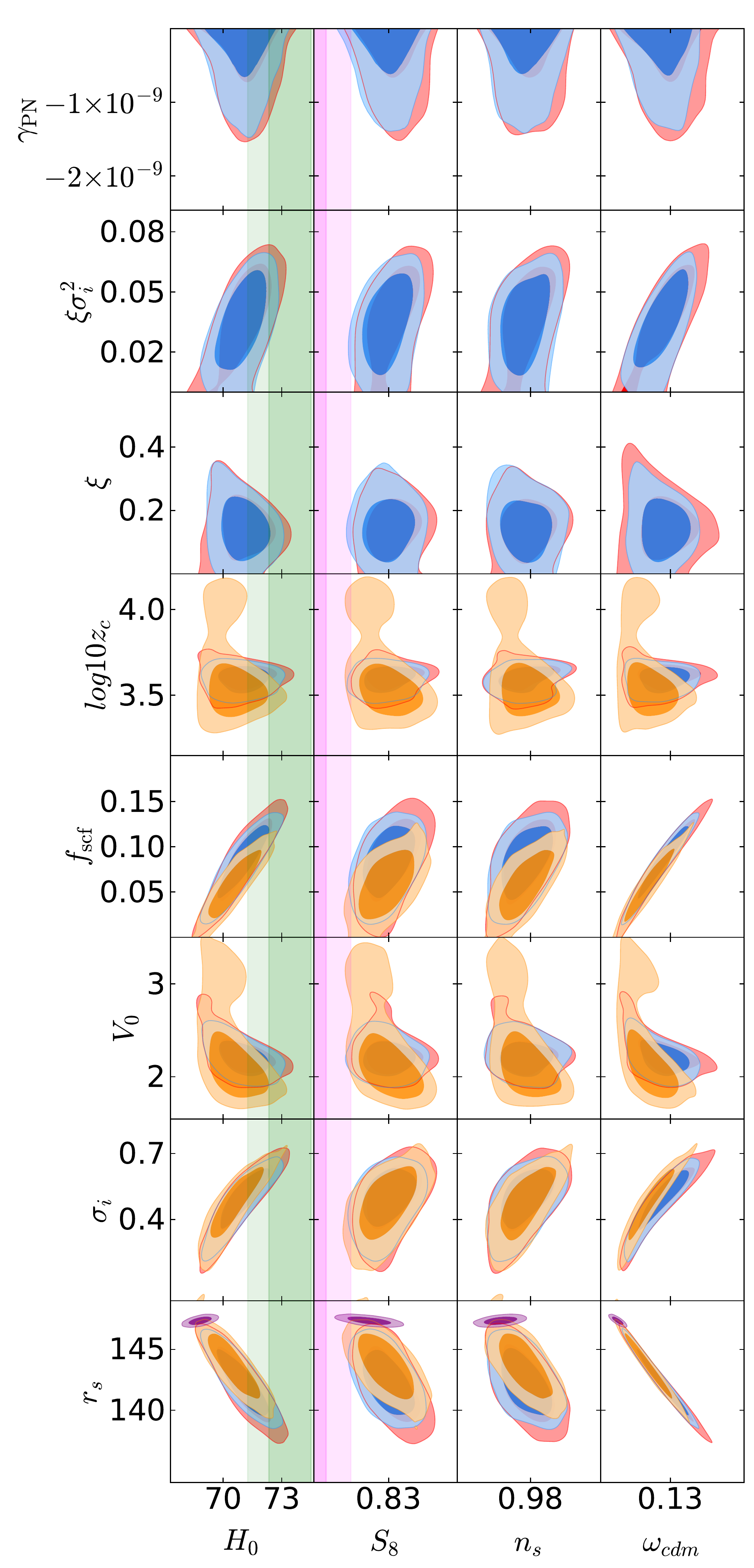}	
		\end{center}
		\caption{\label{fig:P18FSBAOSNH0} 1D and 2D posterior distributions of a subset of parameters for $\Lambda$CDM, EDE and EMG obtained using the data set {\bf P18 + BAO + FS + SN + $H_0$}. Red contours show the results obtained for EMG with a larger prior on $V_0$ (see main text), for which we use the data set {\bf P18 + BAO + SN + $H_0$}. 2D contours contain 68\% and 95\% of the probability. We also plot the 68\% and 95\% CL for the priors on $H_0$ and $S_8$ described in Sec.~\ref{sec:results}. }
	\end{figure}

We start by discussing the results obtained using the data set {\bf P18 + BAO + FS + SN + $H_0$}, which are presented in Fig.~\ref{fig:P18FSBAOSNH0} and Table~\ref{tab:P18BAOFSSNH0}.
They clearly show that in the EMG model a large value of $H_0=71.00^{+0.87}_{-0.79}$  km s$^{-1}$Mpc$^{-1}$ at 68\% CL is obtained, reducing the tension with SH0ES + H0LiCOW at 1.7$\sigma$, better than the 2.1 (4) $\sigma$ reduction for the EDE ($\Lambda$CDM) obtained for $H_0=70.57^{+0.77}_{-0.98}$  $(68.82\pm0.39)$  km s$^{-1}$Mpc$^{-1}$ at 68\% CL. This reduction comes both from the larger mean value of $H_0$ and the larger errors compared to $\Lambda$CDM. As for other models aiming at solving the $H_0$, we obtain a larger $\omega_c$ and  $n_s$ compared to the $\Lambda$CDM model\footnote{Note that, taken at face value, a larger $n_s$ would have profound implications on inflationary physics \cite{Akrami:2018odb,Martin:2013tda,Vennin:2015vfa}. }.
    
It is interesting to note that 
EMG helps fitting CMB data better with respect to EDE (and also to the $\Lambda$CDM). This is reflected in our 68 \% CL estimate for $\xi = 0.15^{+0.06}_{-0.07}$, its 95 \% CL upper limit $\xi<0.42$, and a best-fit value of $\xi=0.178$. 
We also get $\sigma_i=0.49^{+0.11}_{-0.06}$ at 68\%CL, or equivalently $f_{\rm scf}=0.084^{+0.030}_{-0.021}$. Note however the remarks in Section~\ref{sec:model} about the meaning of $f_{\rm scf}$ in the context of EMG. 
    
    \onecolumngrid

\twocolumngrid

Compared to $\Lambda$CDM, both the EMG and the EDE model exacerbate the tension with measurements of $\sigma_8$ and $S_8$. 
We get consistent results in terms of $\sigma_8$ for EMG and EDE, i.e. $\sigma_8=0.830\pm0.008$ at  68\%CL for EMG and $\sigma_8=0.832^{+0.009}_{-0.011}$ at  68\%CL for EDE. 
However, the larger $\omega_c$ and $H_0$ leads to essentially  the same $S_8=0.829\pm0.011 (\pm 0.13)$  at  68\%CL for EMG (EDE).

Overall, the EMG models fits the data much better than the $\Lambda$CDM model with an improvement of $\Delta\chi^2=-16.0$. Such an improvement (better than $\Delta\chi^2=-9.3$ for the EDE model) is largely due to the better fit to the $H_0$ prior, but there is also some improvement in the fit to CMB data, in particular to high-$\ell$ TTTEEE data. As for LSS data, there is only a very small degradation compared to $\Lambda$CDM due to the $\Delta\chi^2=+2.5$ in the fit to BAO DR12 FS + BAO, high-$z$ NGC. The suppression of the matter power spectrum given by the large positive coupling $\xi$ helps fit FS + BAO data keeping the value of $H_0$ large at the same time.  This large improvement in the fit 
corresponds to a Bayes factor of $\ln B_{ij}=+1.0$ for EMG. 
The EDE model, which leads to a smaller improvement in the fit , i.e.   $\Delta\chi^2=-9.3$, has nevertheless  a slightly larger Bayes factor of $\ln B_{ij}=+1.5$ due to the smaller number of extra parameters compared to EMG. Note that, from its definition in Eq.~\eqref{eq:bayesfactor}, the Bayes factor depends on the prior range of the extra parameter $\xi$ and as such has to be interpreted with some caution. In fact, especially if a parameter is not well constrained (as for the case of some the EMG parameters as $V_0$ and $\xi$, see next Section) one could enhance the evidence for the EMG model by reducing the prior range and therefore the sampling volume. For attempts towards  model selection techniques which are less dependent on the specific choice of the prior see e.g. Ref.~\cite{Gariazzo:2019xhx}.

\onecolumngrid

		\begin{figure}
		\begin{center} 
			\includegraphics[width=.7\columnwidth]{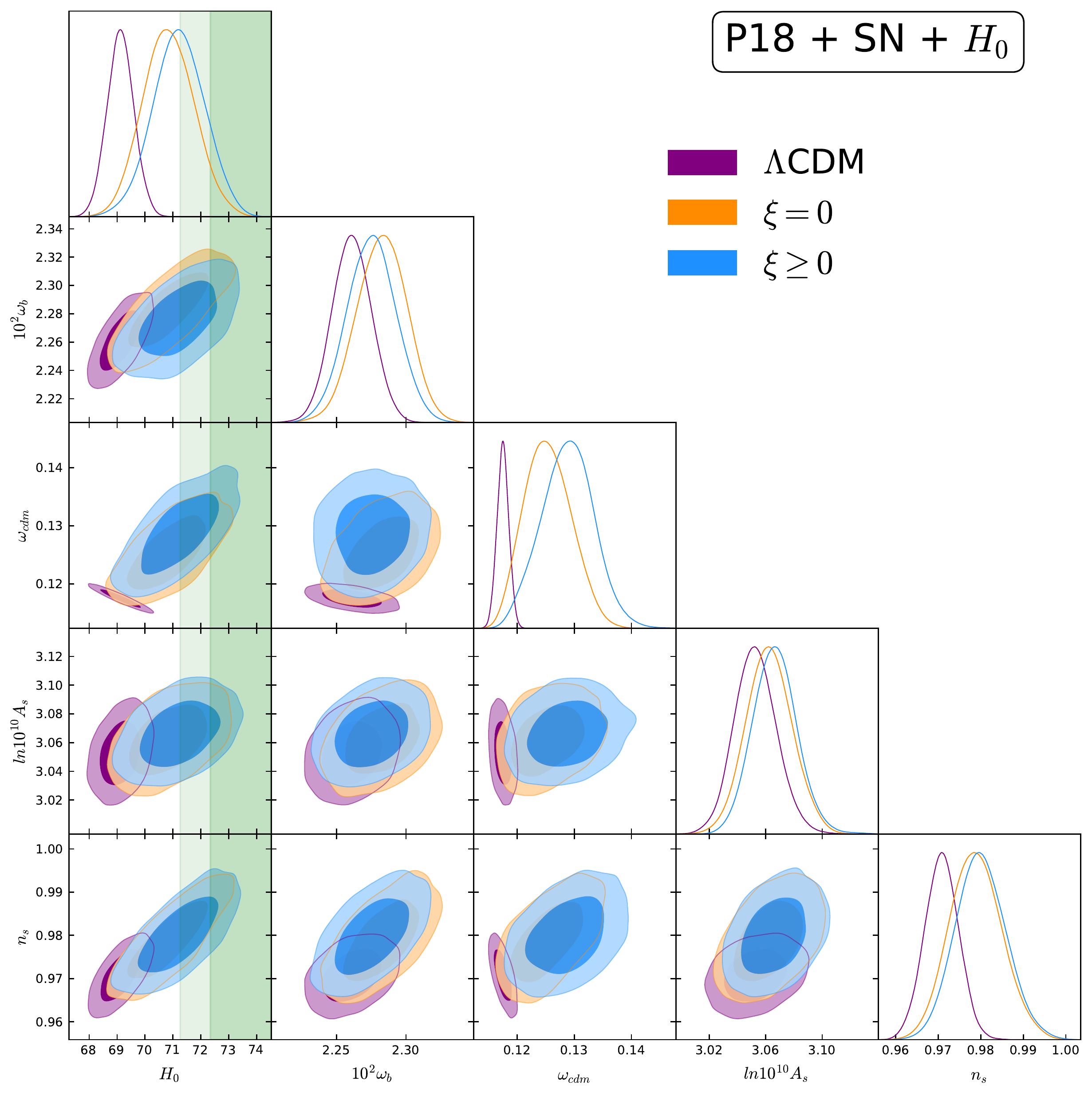}	
			\includegraphics[width=.29\columnwidth]{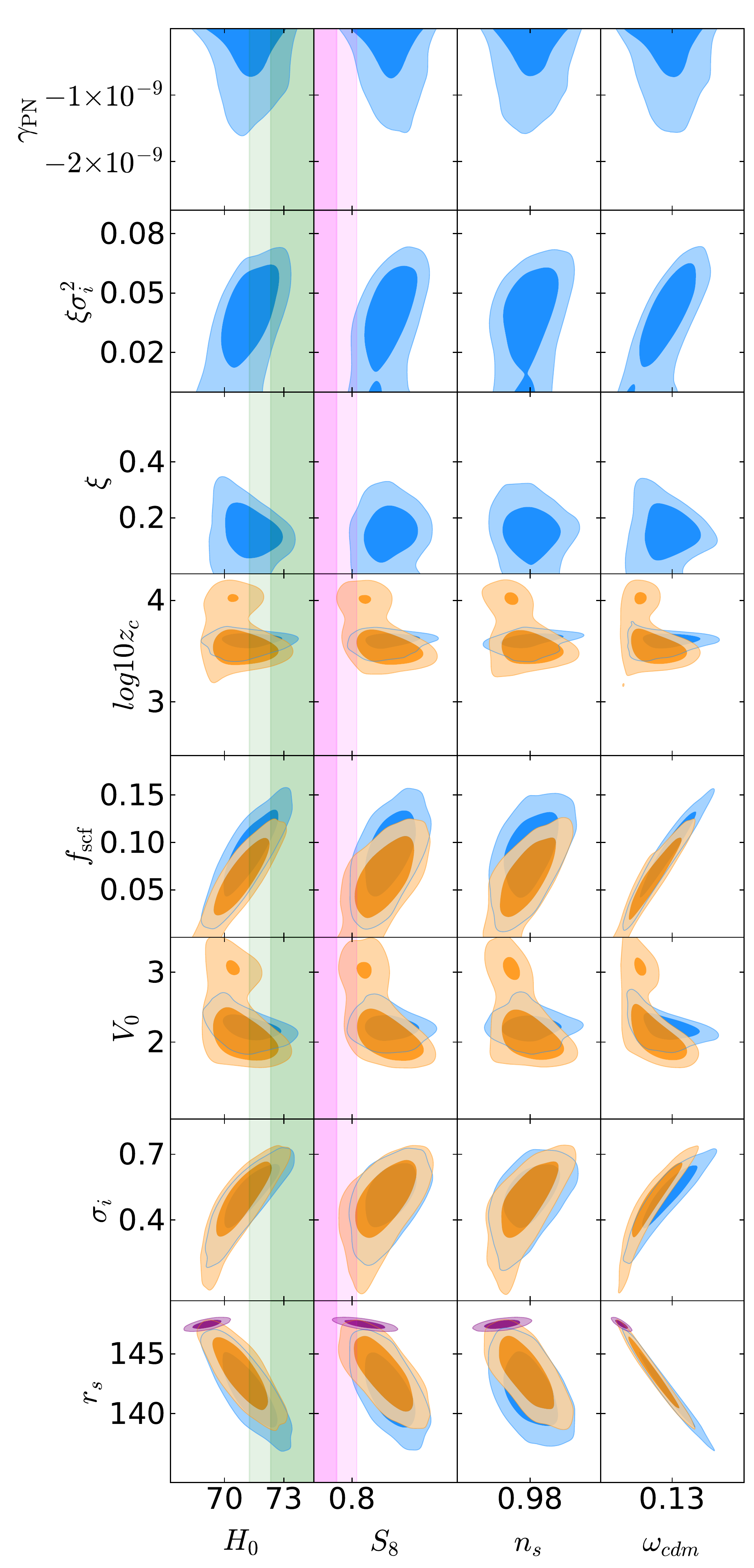}	
		\end{center}
		\caption{\label{fig:P18SNH0}  1D and 2D posterior distributions of a subset of parameters for $\Lambda$CDM, EDE and EMG obtained using the data set {\bf P18 + SN + $H_0$ }. 2D contours contain 68\% and 95\% of the probability. We also plot the 68\% and 95\% CL for the priors on $H_0$ and $S_8$ described in Sec.~\ref{sec:results}. }
	\end{figure}

\twocolumngrid

\hfill\break

With the choice of $V_0$ prior as above, however, it is not possible to recover the model studied in Ref.~\cite{Braglia:2020iik} as the particular $\lambda\to0$ limit. The reason of this choice is to make sure that for every possible combination of parameters the scalar field always decreases toward $\sigma=0$, so to be able to safely use the FS data. Indeed, for $\lambda=0$, the deviation from GR grows at late times, invalidating the use of the FS likelihood and PyBird for a large portion of the parameter space.
    
On the other hand, it is instructive to study the effects of widening the $V_0$ prior to see if the data constrain the model with $\lambda=0$. For this purpose we perform an MCMC analysis with the data set {\bf P18 + BAO  + SN + $H_0$} that does not suffer from the issue raised above and we set the prior range $V_0\in[-4,\,3.5]$. We have checked that for $V_0\leq-3$, the potential is essentially negligible. 
    
The posteriors obtained for this MCMCs analysis are shown as red contours in Fig.~\ref{fig:P18FSBAOSNH0} and they show that data do not prefer the small $V_0$ region for which the scalar field grows. The results also show another interesting feature of the EMG model, i.e. there is only a small difference in constraints on the EMG model when using BAO in place of the more complete BAO + FS data. As can be seen, the only effect of using BAO is have slightly larger posteriors, but with the same mean as those obtained with BAO + FS data. Note that this is in agreement for the findings of Ref.\cite{Niedermann:2020qbw} in the context of the New Early Dark Energy model.

\onecolumngrid

		\begin{figure}
		\begin{center} 
			\includegraphics[width=.7\columnwidth]{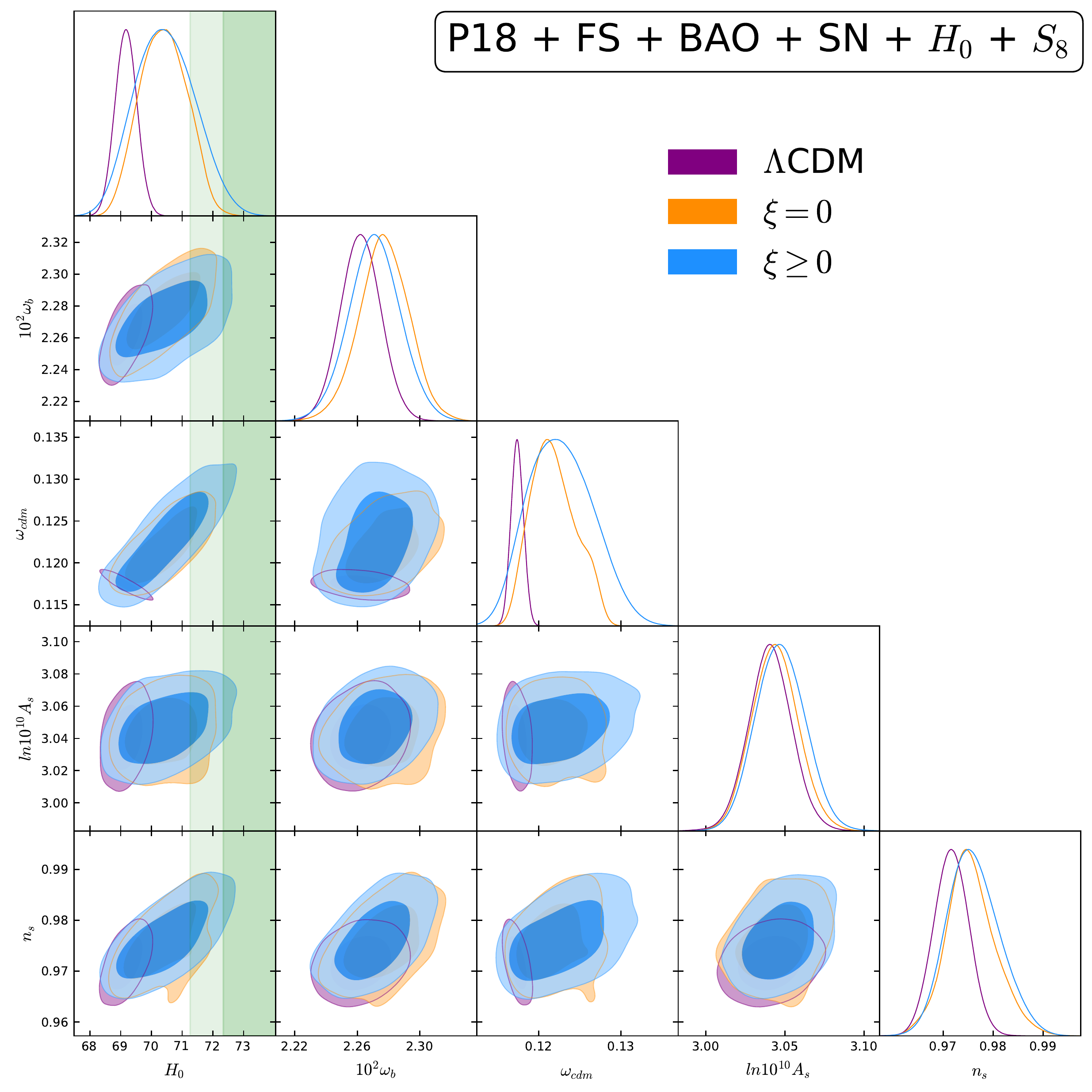}
			\includegraphics[width=.29\columnwidth]{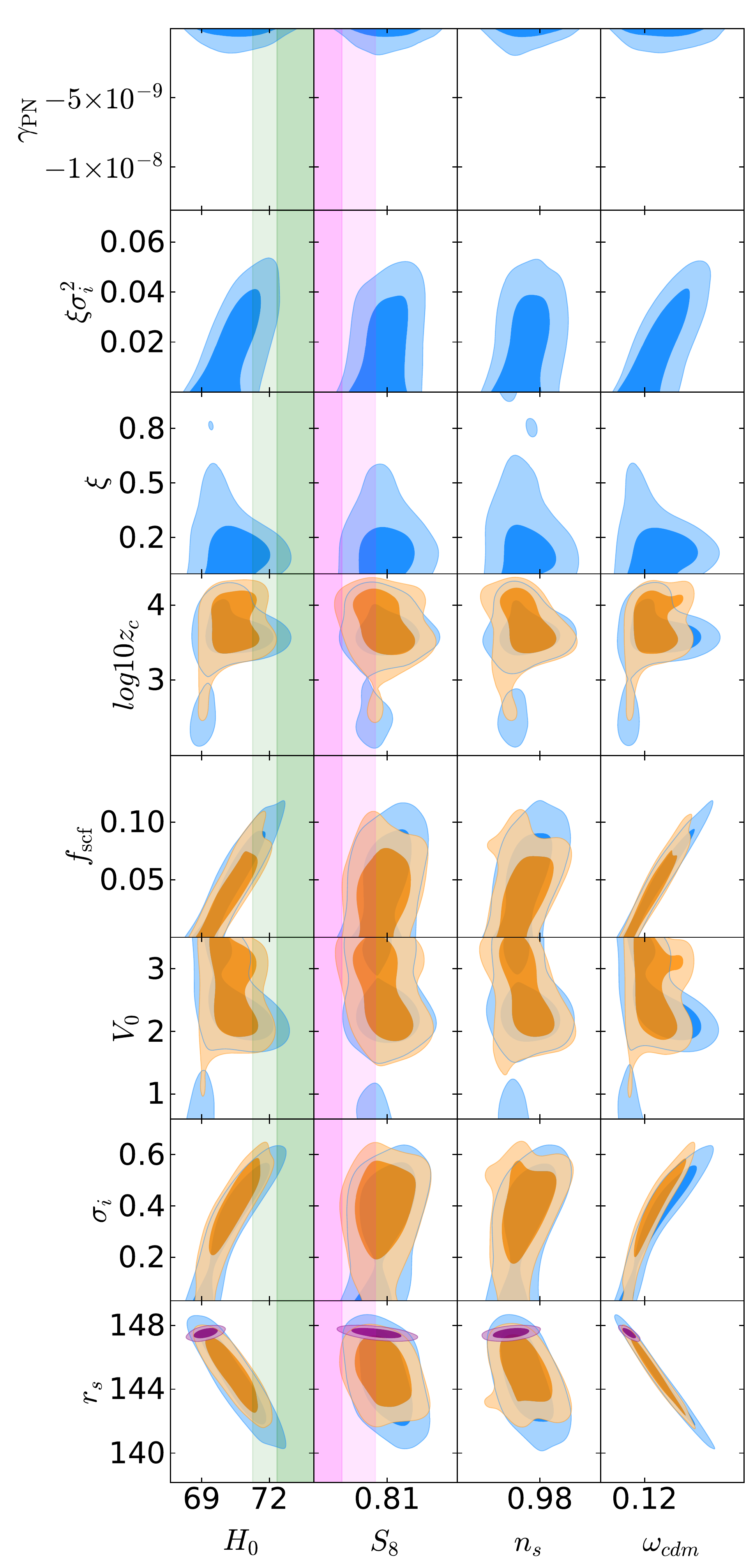}	
		\end{center}
		\caption{\label{fig:P18FSBAOSNH0S8}  1D and 2D posterior distributions of a subset of parameters for $\Lambda$CDM, EDE and EMG obtained using the data set {\bf P18 + BAO + FS + SN + $H_0$ + $S_8$}. 2D contours contain 68\% and 95\% of the probability. We also plot the 68\% and 95\% CL for the priors on $H_0$ and $S_8$ described in Sec.~\ref{sec:results}. }
	\end{figure}

\twocolumngrid

\hfill\break

In order to further assess the role of BAO + FS data, we also perform  an MCMC analysis without considering them, and use the data set {\bf P18 + SN + $H_0$}. The results are presented in Fig.~\ref{fig:P18SNH0} and Table~\ref{tab:P18SNH0}. As can be seen, removing BAO and FS data leads to a somewhat larger value of $H_0 = 70.85\pm0.92 $ km s$^{-1}$Mpc$^{-1}$ for the EDE (and a much larger bestfit of $H_0 = 71.38$ km s$^{-1}$Mpc$^{-1}$), confirming that BAO + FS have the power to constrain these models, as shown in\footnote{Note that our  results slightly differ from the ones in Ref.~\cite{DAmico:2020ods}. The main reason is that we use a different  $H_0$ prior which has a stronger impact on our MCMC analysis, whereas they used a prior obtained from earlier SH0ES results, i.e. $H_0 = 74.03 \pm 1.42$ km s$^{-1}$Mpc$^{-1}$ \cite{Riess:2019cxk}. Also, we fix $N_{\rm eff}=3.046$ in our analysis which leads results slightly different from the ones obtained with the Planck assumption of one massive neutrino with $m_\nu=0.06$ eV (see main text).  Finally, we do not use high redshift Lyman-$\alpha$ forest data from eBOSS DR14 measurements \cite{Agathe:2019vsu,Blomqvist:2019rah} . We have checked that we recover the results that are consistent with Ref.~\cite{DAmico:2020ods} when using the EDE model with $\xi=0$ and their dataset and conventions.} Refs.~\cite{Hill:2020osr,Ivanov:2020ril,DAmico:2020ods}. On the other hand, $H_0$ for the EMG model increases only a bit to $H_0=71.21\pm0.93$ km s$^{-1}$Mpc$^{-1}$, since  BAO + FS data constrain it less than they constrain EDE models. It is very interesting to note that the best-fit value for the coupling $\xi=0.17$ is very close to the one found including BAO + FS data.

\onecolumngrid

		\begin{figure}
		\begin{center} 
			\includegraphics[width=.7\columnwidth]{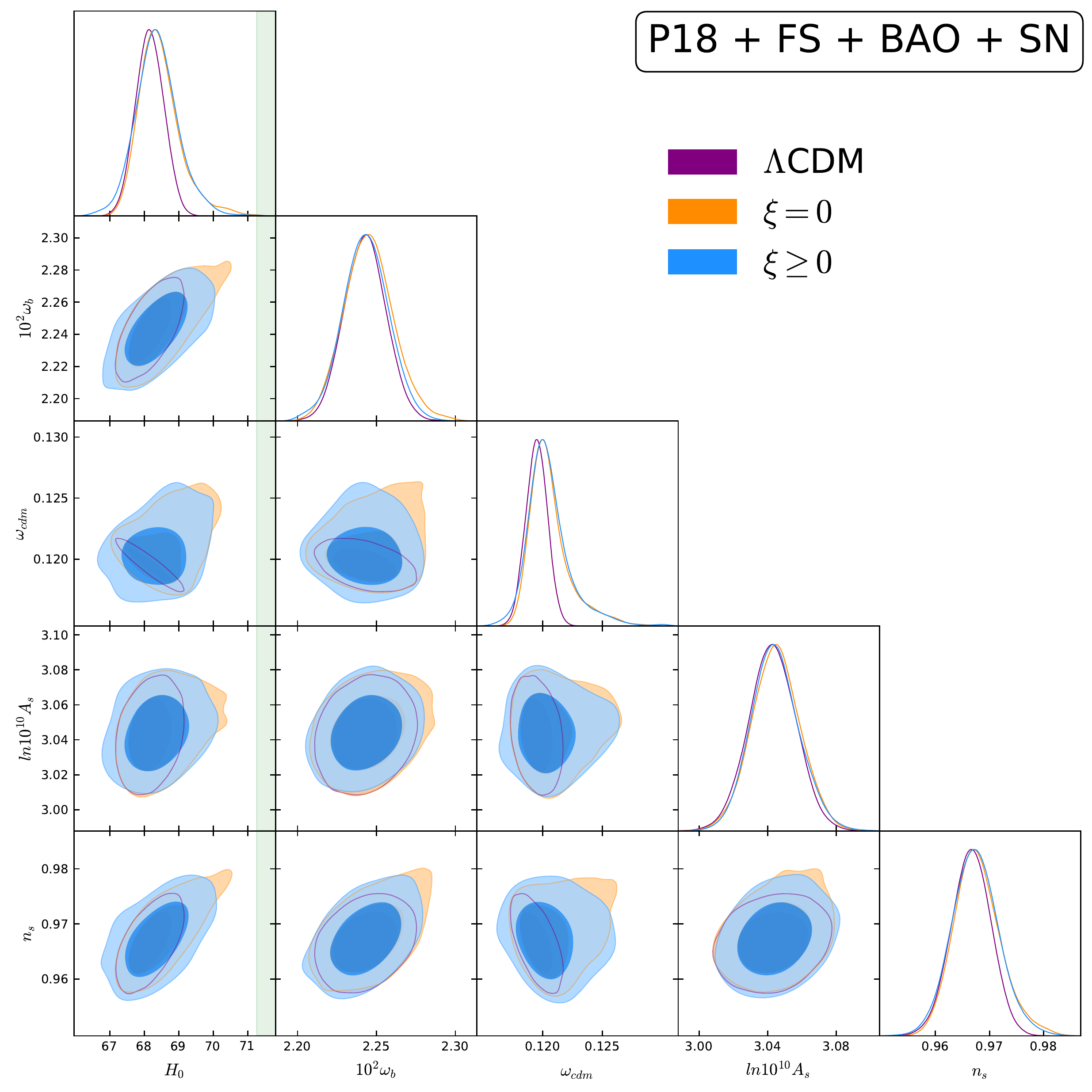}	
			\includegraphics[width=.29\columnwidth]{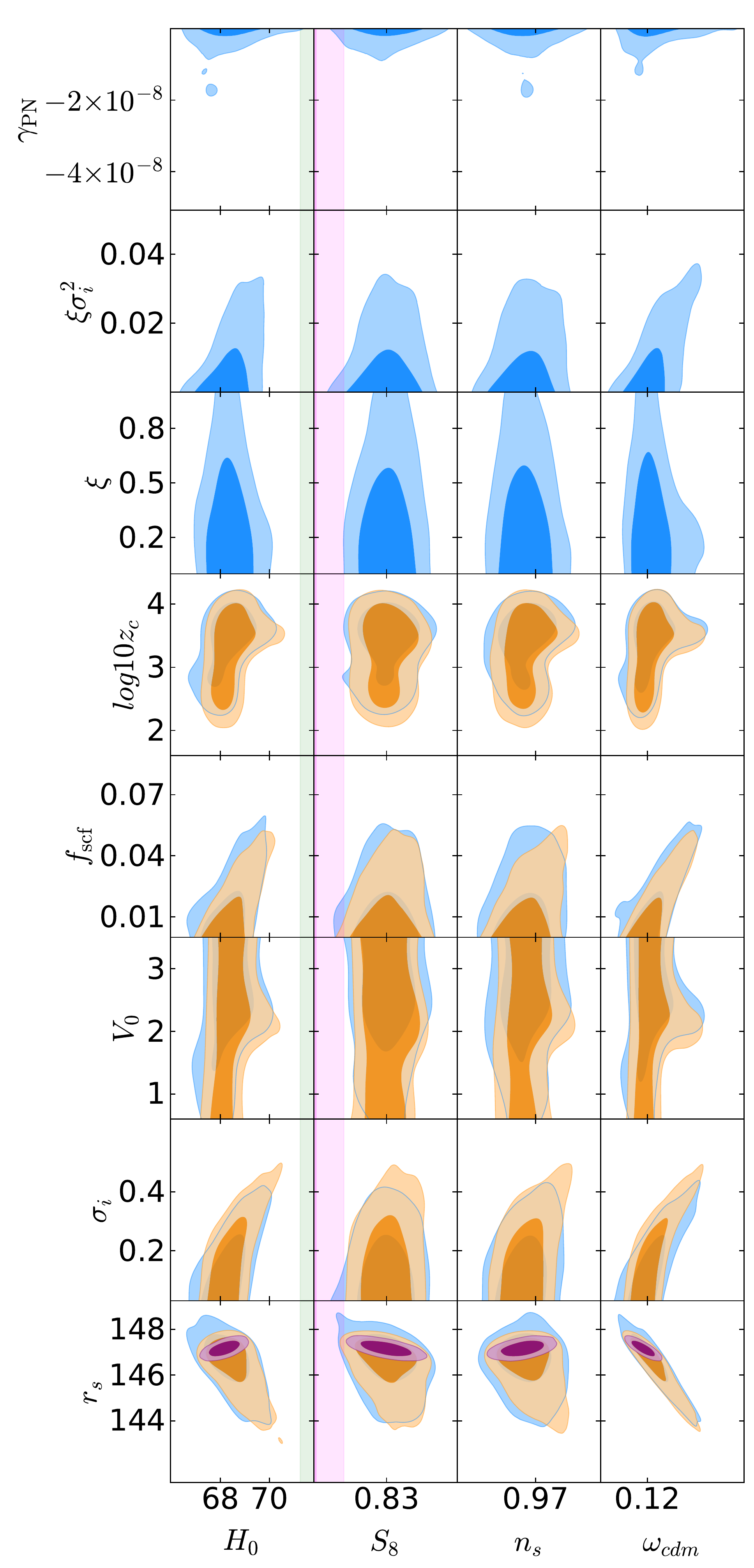}	
		\end{center}
		\caption{\label{fig:P18FSBAOSN} 1D and 2D posterior distributions of a subset of parameters for $\Lambda$CDM, EDE and EMG obtained using the data set {\bf P18 + BAO + FS + SN }. 2D contours contain 68\% and 95\% of the probability. We also plot the 68\% and 95\% CL for the priors on $H_0$ and $S_8$ described in Sec.~\ref{sec:results}. }
	\end{figure}

\twocolumngrid

The EMG model fits most of the data, with the exception of CMB lensing, better than both the EDE and the $\Lambda$CDM model, leading to a $\Delta \chi^2=-17.1$. This time, however, the improvement in the fit  does not warrant the increase in the model complexity compared to $\Lambda$CDM and we obtain a Bayes factor of $\ln B_{ij}=-0.2$.
    
We have shown that the EMG model leads to a larger value of $S_8$ compared to the $\Lambda$CDM one. Therefore, we would like to test it against weak lensing data. Strictly speaking, this would require using data from e.g. the KiDS-VIKING galaxy shear measurements. However, it was claimed in Refs.~\cite{Hill:2020osr,Ivanov:2020ril} that the same results can be obtained by implementing weak lensing data through a Gaussian prior on the parameter $S_8=0.770\pm0.017$ (see also Ref.~\cite{Murgia:2020ryi} for a thorough comparison of this method to the correct use of cosmic shear measurements). With these caveats, we follow Refs.~\cite{Hill:2020osr,Ivanov:2020ril} and present in Fig.~\ref{fig:P18FSBAOSNH0S8} and Table~\ref{tab:P18BAOFSSNH0S8} the results for the data set {\bf P18 + BAO + FS + SN + $S_8$ + $H_0$}. Note, despite being far from a resolution to the $S_8$ tension, the EMG model shows now a much smaller $S_8=0.809\pm0.009$ and a bestfit value of $S_8=0.807$, lower than the one obtained for $\Lambda$CDM i.e. $S_8=0.811$. This confirms the conclusion of Ref.~\cite{Murgia:2020ryi} for EDE models that, even though it is true that the $S_8$ tension is not resolved within this model, the same holds for the $\Lambda$CDM model which, however, is not able to address the $H_0$ tension, as opposed to the EMG model, for which we obtain a mean $H_0 = 70.63^{+0.80}_{-1.00} $ and a best fit of $H_0=71.59$ km s$^{-1}$Mpc$^{-1}$.

Even in this case, however, we note that the large improvement in the fit (not followed by a preference from the model-selection point of view) is coming mainly from the substantial improvement in the fit to $H_0$. It is therefore natural to ask what happens when we remove $H_0$ prior from the data set. 

We present the results obtained 
without 
the combined SH0ES-H0LiCOW determination of $H_0$ in Fig.~\ref{fig:P18FSBAOSN} and Table~\ref{tab:P18BAOFSSN} for the data set {\bf P18 + BAO + FS + SN}\footnote{We have checked that the addition of a prior on $S_8$ to this data set does not change appreciably the results. }. The results show that the mean value for $H_0$ in the EMG model (and in the EDE one) is only slightly larger then the one in $\Lambda$CDM, as also found in previous studies of effectively massless models of scalar-tensor theories \cite{Umilta:2015cta,Ballardini:2016cvy} .
This can be appreciated by looking at the larger posterior distributions of $H_0$ and $\omega_c$ for the EMG and EDE models in Figs.~\ref{fig:P18FSBAOSN}. The incapability of EDE to solve the $H_0$ tension when prior information on $H_0$ is not included, has been recently discussed in the literature \cite{Hill:2020osr}. 
A similar result holds for EMG. \footnote{However, it has also been proposed in Refs.~\cite{Murgia:2020ryi,Smith:2020rxx} (see also next Section), that a distinction should be made between looking at the posterior distributions and the fact that there are some parameters that fit the data in a way that is statistically indistinguishable from $\Lambda$CDM and still lead to a large $H_0$.}

Although the best-fit parameters shown in the third column of Table~\ref{tab:P18BAOFSSN} do not lead to a very large $H_0$, we confirm the results of Refs.~\cite{Murgia:2020ryi,Smith:2020rxx} for EMG and find some set of parameters exist that lead to a large $H_0$ without a significant change in $\Delta\chi^2$. For example, we find that  $100\,\omega_b=2.285,\,$
 $\omega_c=0.1308,\,$ $100*\theta_s=1.04089,\,$ $\tau_{\rm reio}=0.057,\,$ $\ln 10^{10}A_s= 3.066\,$ $n_s=0.9840,\,$ $\xi=0.151,\,$ $V_0=2.19,\,$ and $\sigma_i=0.57$ leads to  $\Delta \chi^2=0.7$, fitting the data very similarly to $\Lambda$CDM, with an improvement in the fit to CMB data and a slight worsening to the fit to BAO DR12 FS + BAO, high-$z$ NGC data. Such a parameter set, leads to a large $f_{\rm scf}=0.081$ and and a large $H_0=70.15$ km s$^{-1}$Mpc$^{-1}$.

\section{Analysis of the 1 parameter extension}
\label{sec:1par}

The $\Lambda$CDM model predictions can be recovered in both the EDE and the EMG models when $\sigma_i$, or equivalently the energy injection of the scalar field into the cosmic fluid, goes to zero. In this regime, both $V_0$ and the coupling $\xi$ essentially play no role. When using the Metropolis-Hasting algorithm, this can give rise to a large portion of the parameter space that can artificially enhance the statistical weight of $\Lambda$CDM models. This issue has been recently addressed, within EDE models, in Refs.~\cite{Smith:2019ihp,Lin:2019qug,Niedermann:2019olb,Niedermann:2020dwg,DAmico:2020ods}. 

Here, we take a similar, but somewhat alternative approach, and follow the  lines of Refs.~\cite{Murgia:2020ryi,Smith:2020rxx}, where it was shown that by fixing\footnote{In the EDE model of Refs.~\cite{Murgia:2020ryi,Smith:2020rxx} also a second parameter related to the axion decay constant $f$, namely $\Theta_i$, has to be fixed.} $\log_{10}z_c$ (or $V_0$ in our convention) it is possible to extend the $f_{\rm scf}-H_0$ degeneracy even for a choice of datasets without prior information on $H_0$, avoiding problems related to the volume sampling and to the choice of a prior that allows for a $\Lambda$CDM limit. Such a degeneracy is clearly disrupted (see Fig.~\ref{fig:P18FSBAOSN}) when a prior on $H_0$ is not included in the data set and a tight upper bound on $f_{\rm scf}$ is obtained. 

Note, however, that in absence of theoretical motivations, this must be seen only as a purely phenomenological approach, which is rather unorthodox from the standard Bayesian point of view, for which all the parameters has to be varied altogether. 
Nevertheless, in the class of MG considered here, there is however the possibility to reduce the number of parameters by restricting to $\xi=-1/6$, which corresponds to the theoretical value for conformal coupling \cite{Rossi:2019lgt} (see more in the following Section).

\begin{figure}
		\begin{center} 
			\includegraphics[width=.7\columnwidth]{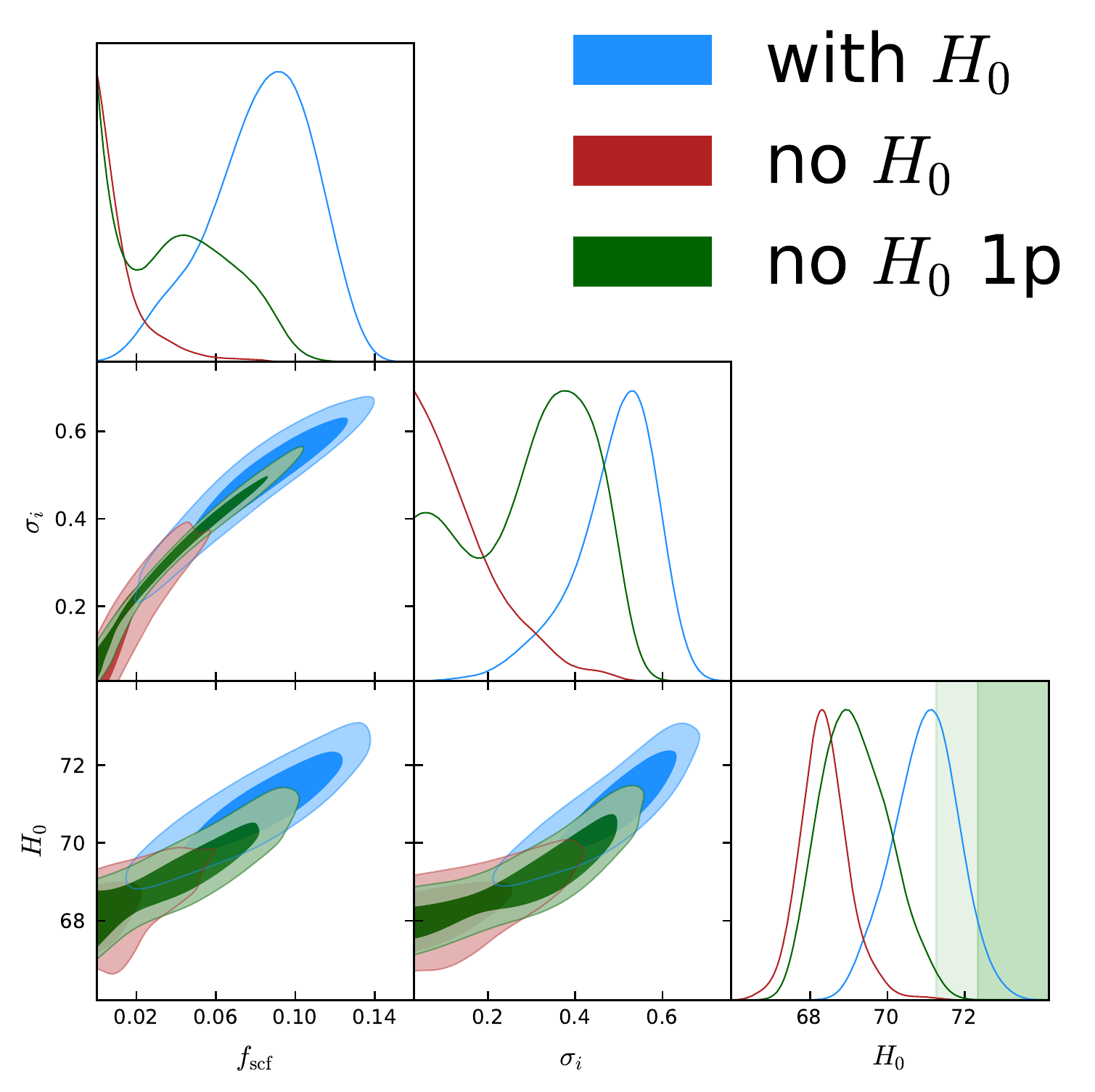}
		\end{center}
		\caption{\label{fig:rpvs1p} 1D and 2D posterior distributions of a subset of parameters for the EMG model.  Blue and dark red contours are drawn from the samples in Fig.~\ref{fig:P18FSBAOSNH0} and    \ref{fig:P18FSBAOSN} respectively, while the green ones are drawn from the sample obtained by using the data set {\bf P18 + BAO + FS + SN} and fixing $V_0$ and $\xi$ to their best-fit values in the third column of Table~\ref{tab:P18BAOFSSNH0}. 2D contours contain 68\% and 95\% of the probability. We also plot the 68\% and 95\% CL for the prior on $H_0$ described in Sec.~\ref{sec:results}. }
	\end{figure}

Based on the former argument, we perform an analysis similar to the one of Ref.~\cite{Niedermann:2020dwg,Murgia:2020ryi} for the EMG model, for which we fix $V_0$ and $\xi$ to their best-fit values in the third column of Table~\ref{tab:P18BAOFSSNH0} and leave $\sigma_i$ free to vary. We do not include $H_0$ data and we use the {\bf P18 + BAO + FS + SN} data set. The results are presented in Fig.~\ref{fig:rpvs1p}, where we confront our results to ones for EMG obtained in the previous section considering the data sets {\bf P18 + BAO + FS + SN} and {\bf P18 + BAO + FS + SN + $H_0$}.

From the plot, it is easy to see that the degeneracy between $\sigma_i$ and $H_0$ is now more visible leading to a larger of $H_0=69.18^{+0.79}_{-1.00}$  km s$^{-1}$Mpc$^{-1}$  at 68\% CL and slightly reducing the tension with SH$0$ES+H0LiCOW ($3.2 \sigma$ vs $4.2 \sigma$ in the 3 parameter case using the same data set). However, the value of $\sigma_i$ remains consistent with $\sigma_i=0$ and most of the improvement in reducing the tension is ascribed to a larger error on $H_0$ compared to the 3 parameters case. In fact, the best-fit value for $H_0$ that we obtain is $H_0=68.79$  km s$^{-1}$Mpc$^{-1}$, corresponding to $\sigma_i=0.30\,M_{\rm pl}$. The best-fit cosmology for the 1 parameter EMG leads to a total $\chi^2$ of $4001.5$, i.e. $\Delta\chi^2=1.8$, nearly indistinguishable from the 3 parameters one. Compared to the 3 parameters models we have a $\Delta\chi^2\sim-1.3$ and a $\Delta\chi^2\sim -0.9$ gain in the fitting Planck high-$\ell$ TTTEEE data and  BAO DR12 FS + BAO, low-$z$ NGC and high-$z$ SGC
respectively, whereas the fit to BAO DR12 FS + BAO, high-$z$ NGC is worsen by a factor of $\sim+1.6$, all the other partial $\chi^2$s being essentially the same.

It is interesting to note that now there is only $1$ extra parameter and the model is not as penalized as for the case with 3 parameters. In fact, the Bayes factor is now $\ln B=1.4$ and for the data set {\bf P18 + BAO + FS + SN}, the model results slightly preferred over $\Lambda$CDM according to the Jeffreys scale in Table~\ref{tab:kassraftery}. 

We therefore conclude that by fixing two parameters does not help much alleviate the $H_0$ tension, which is only addressed when additional prior information from local measurements of the Hubble constant is added, as shown in the previous Section. As in Section~\ref{sec:results}, though, we note that we do find some choices of parameters  for which the fit to the data is not substantially different from the one in the $\Lambda$CDM model, but lead to a larger $H_0$, as in Refs.~\cite{Murgia:2020ryi,Smith:2020rxx}, with which we qualitatively agree. A fully quantitative comparison with Refs.~\cite{Murgia:2020ryi,Smith:2020rxx} is however not possible because of the presence of the non-minimal coupling and the different potential considered. 
\onecolumngrid

\twocolumngrid	

\hfill \break
Indeed, potentials with a different curvature such as those with flattened wings and power-law minima are well known to lead to a larger value of $H_0$ compared to the simpler quartic potential \cite{Poulin:2018cxd,Lin:2019qug,Smith:2019ihp,Braglia:2020bym}.

As a last comment, note that it would be an interesting exercise to fix also $\xi$ to search for the $\sigma_i$ that exactly solves the $H_0$ tension, as proposed in Ref.~\cite{Vagnozzi:2019ezj}. However, this is not the purpose of this paper and we hope to return to this point in a future work.

\section{The $\xi=-1/6$ case}
\label{sec:conf}
So far, we have focused on the case of a positive coupling $\xi\geq0$. As a representative example of the parameter space with $\xi<0$, we also show the results obtained by fixing $\xi=-1/6$ \cite{Rossi:2019lgt}. 

From Fig.~\ref{fig:Background} in Section~\ref{sec:model}, we see that the energy injection is not sharp in redshift anymore, but rather we observe a continuous energy injection in the early Universe, until the scalar field contribution redshits away.

The similarity between the background dynamics of this model and the one of a model with extra dark-radiation parameterized by $N_{\rm eff}$ and the consequent difficulty in constraining the coupling $\xi$ has been studied in Ref.~\cite{Braglia:2020iik}. Here, the contribution of the scalar field to the total energy budget is similar so we do not expect significant differences between the results here and the ones found in Ref.~\cite{Braglia:2020iik}. However, note that thanks to the small effective mass, the scalar field decreases more rapidly compared to the massless case with $\lambda=0$, see e.g. Fig.~1 of Ref.~\cite{Braglia:2020iik}.

\onecolumngrid

		\begin{figure}[h!]
		\begin{center}
			\includegraphics[width=.67\columnwidth]{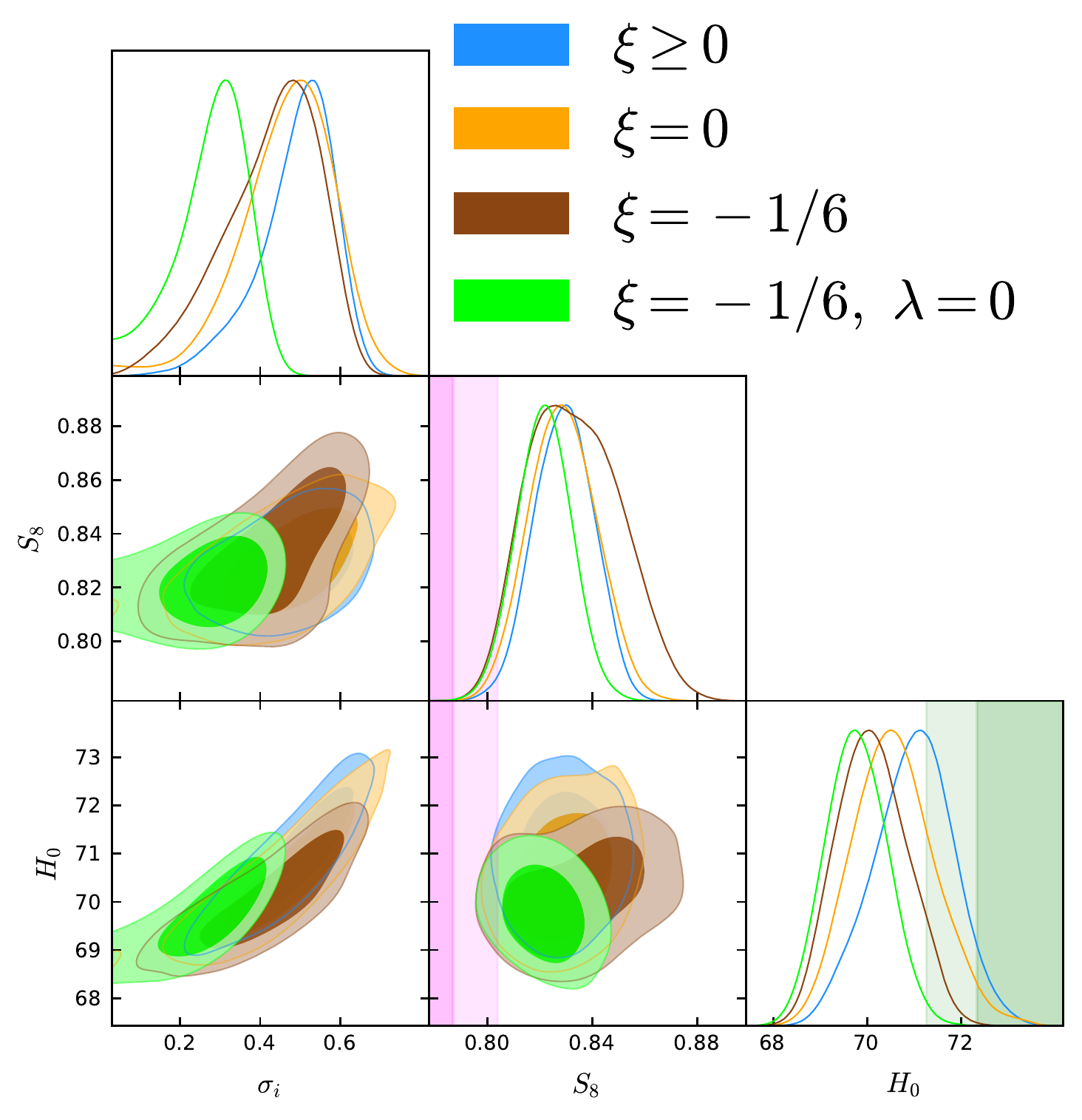}
			\includegraphics[width=.32\columnwidth]{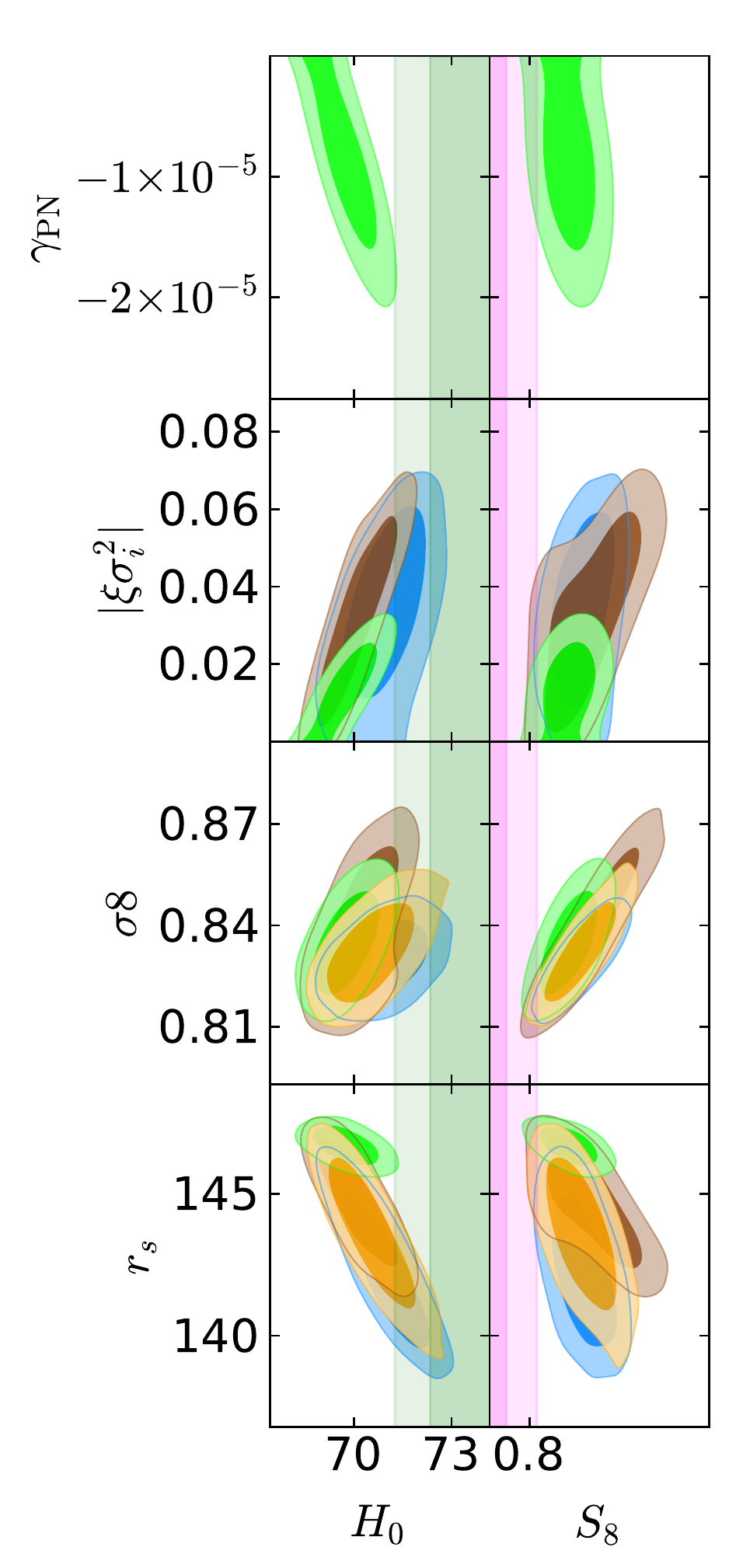}
		\end{center}
		\caption{\label{fig:Conf}  1D and 2D posterior distributions of a subset of parameters for the EDE (orange), EMG (blue) and conformally coupled EMG (brown) using the data set {\bf P18 + BAO + FS + SN + $H_0$}. We also show in green the results for the case with $\xi=-1/6$ and $\lambda=0$ for a comparison. Note that for the latter case the data set {\bf P18 + BAO  + SN + $H_0$} is instead used. 2D contours contain 68\% and 95\% of the probability. We also plot the 68\% and 95\% CL for the priors on $H_0$ and $S_8$ described in Sec.~\ref{sec:results}. }
	\end{figure}

\twocolumngrid

\hfill\break
For our MCMC analysis we use the data set {\bf P18 + BAO + FS + SN + $H_0$} and we fix $\xi=-1/6$. 
Our results are shown in Fig.~\ref{fig:Conf}, where we compare to results of the previous section and show also the results for the case with $\xi=-1/6$ and $\lambda=0$ obtained with the same prior on $\sigma_i$ for a comparison (for simplicity we refer to it as CC). Note that, for the $\lambda=0$, we have used the data set {\bf P18 + BAO + SN + $H_0$}, since for a large portion of the $\sigma_i$ prior we have $G_{\rm eff}/G-1\sim 10^{-3}$ and the use of the FS likelihood might be less  accurate.

Fig.~\ref{fig:Conf} shows that the EMG case with $\xi=-1/6$ leads to $H_0=70.11\pm 0.79$  km s$^{-1}$Mpc$^{-1}$ at 68\% CL
a value smaller than the one obtained in the EDE and EMG model with $\xi\geq0$.  This is expected, as the ability of the EDE and  EMG model with $\xi\geq0$ to alleviate the $H_0$ tension relies on an energy injection very localized in redshift, a feature that is not shared by the EMG model with $\xi=-1/6$. The bestfit value of $\sigma_i=0.46\, M_{\rm pl}$ leads to $H_0=70.30$  km s$^{-1}$Mpc$^{-1}$,  again smaller than the $\xi=0$ and $\xi\leq0$ case. The improvement in the fit is $\Delta\chi^2=-9.0$ accompanied by a Bayes factor of $\ln B_{ij}=-1.4$, as in the EDE case, which has the same number of parameters. The main improvement in the $\Delta\chi^2$ comes from a better fit to Planck high-$\ell$ data compared to the other EMG and EDE models, but it is compensated by a degradation in the fit to LSS and $H_0$ data.

On the other hand, in the latter model, the energy density of the scalar field redshifts away much faster than for $\lambda=0$, since the scalar field is driven towards $\sigma\simeq0$ by the quartic potential. This is the reason why the $H_0$ in this model is larger than $H_0= 69.78\pm 0.66 $  km s$^{-1}$Mpc$^{-1}$ at 68\% CL, obtained for $\lambda=0$, for which the scalar field contribution is not completely negligible after recombination. For the very same argument, we observe that a larger $\lvert\xi\sigma_i^2\rvert$, which is a measure of the scalar field contribution to the fractional $\Delta H(z)/H(z)$ before recombination when $\xi<0$, is allowed in the EMG model compared to the CC one. Also, the value of $\gamma_{\rm PN}$ is orders of magnitude larger in the CC model, i.e. $\gamma_{\rm PN}> -2.1\cdot 10^{-5}$ at 95\%CL, compared to the EMG case with $\xi=-1/6$ in which $\gamma_{\rm PN}> -3.5\cdot 10^{-9}$
at 95\%CL. If the former is comparable to Solar System experiments, the latter is much smaller.

Furthermore, as expected from the discussion in Section~\ref{sec:model} and Fig.~\ref{fig:Pk}, the negative coupling leads to larger $\sigma_8$.   We get $\sigma_8 = 0.837^{+0.013}_{-0.021}$  and $\sigma_8 = 0.835\pm 0.010$ for $\lambda \ne 0$ and $\lambda=0$, respectively,
larger than the EDE or EMG model with a positive coupling (see Table~\ref{tab:P18BAOFSSNH0}). However, this is accompanied by a comparable $S_8 = 0.833^{+0.016}_{-0.022}$ for EMG with $\xi=-1/6$ and a smaller $S_{8} = 0.822\pm 0.011 $ for $\xi=-1/6 \,, \lambda=0$, since $H_0$ is smaller and therefore the shift in the value of $\omega_c$ necessary to restore the fit with CMB data is slightly smaller as well. 
This is again in line with the observation that models that lead to a larger $H_0$ modifying the sound horizon inevitably lead to a larger $\omega_c$ and therefore $S_8$ \cite{Jedamzik:2020zmd}.

\section{Conclusions} \label{sec:conclusion}

We have presented a model of Early Modified Gravity (EMG) where a scalar field with a non-minimal coupling to the Ricci scalar of the type $M^2_{\rm pl}+\xi \sigma^2$ has a self-interacting potential. In this model, which extends the massless one of Ref.~\cite{Braglia:2020iik} and reduces to the Rock'n'Roll Early Dark Energy (EDE) model of Ref.~\cite{Agrawal:2019lmo} for $\xi=0$, the scalar field $\sigma$, which is frozen during radiation era, grows around the time of recombination driven by the coupling to pressureless matter and is subsequently driven into damped oscillations around its minimum at $\sigma=0$ by the small effective mass induced by the quartic potential. The rolling of the field towards $\sigma=0$ suppresses the modification to gravity at late times, recovering an excellent agreement of the laboratory experiments and Solar System tests with General Relativity. The addition of the effective potential has the virtue of reconciling the $\xi > 0$ branch of the model studied in \cite{Braglia:2020iik} with GR without any fine tuning.

The modification to gravity at early times, however, has the important consequence of alleviating the $H_0$ tension as it modifies the redshift profile of the energy injected into the cosmic fluid when the scalar field thaws. Our MCMC analysis, performed with a variety of cosmological data, shows that the tension can be reduced substantially and at the same time a positive coupling $\xi>0$ suppresses the small scale matter power spectrum and thus helps fit the full Shape of the matter power spectrum data, that has recently claimed to constrain the EDE resolution of the $H_0$ tension. In particular, the tension with the combination of recent  SH0ES and H0LiCOW measurements, i.e. $H_0 = 73.4\pm1.1$ km s$^{-1}$Mpc$^{-1}$, is reduced at the $1.7\sigma$ level when this is added to  Cosmic Microwave Background, SNe, Baryonic Acoustic Oscillations and the Full Shape of the matter power spectrum data.  For this data set, we obtain $H_0=71.00^{+0.87}_{-0.79} $ km s$^{-1}$Mpc$^{-1}$ at 68 \% CL which is larger than the one we get for EDE for $\xi=0$ i.e. $H_0=70.57^{+0.77}_{-0.98} $ km s$^{-1}$Mpc$^{-1}$.

Performing the MCMC analysis with different combinations of the data mentioned above helps us trace the origin of the larger $H_0$ back to the suppression of the power spectrum caused by the non minimal coupling $\xi$, for which we get a mean value of $\xi= 0.15^{+0.06}_{-0.07}$ at 68\% CL (however it is only an upper bound $\xi < 0.39$ at 95\%CL). In fact, for all the data set that we use we get a similar constrain on the parameter $\xi$.  Although the fit to data is always improved, however, the Bayesian model selection for EMG depends on the data set considered, and is penalized by the larger number (3) of extra parameters compared to $\Lambda$CDM, therefore never resulting in a strong preference.

In order to confirm the argument above we have performed the same analysis fixing $\xi$ to the conformal coupling $\xi=-1/6$. In this case rather than a suppression we have an enhancement of the matter power spectrum and the capability of the model to ease the tension is therefore reduced, with  $H_0=70.11\pm 0.79  $ km s$^{-1}$Mpc$^{-1}$,  smaller than the $\xi=0$ case, showing a clear hierarchy for negative, null and positive couplings. Note, however, that the addition of the small effective mass to the $\xi=-1/6$ case leads to larger $H_0$ than the one for the conformally coupled massless case of Ref.~\cite{Braglia:2020iik} for which $H_0=69.78\pm 0.66  $ km s$^{-1}$Mpc$^{-1}$ (see Section~\ref{sec:conf}).

As a last comment, in this paper we have considered two dimensionless couplings for a cosmological scalar field, which rule the coupling to the Ricci scalar ($\xi$) and 
its self-interaction ($\lambda$).
A quartic potential for the scalar field $\sigma$, implies that we recover the RnR model \cite{Agrawal:2019lmo} for $\xi=0$. However, it is known that
potentials with flattened wings that have a different curvature around the minimum at $\sigma=0$, such as those in the original EDE proposal of Ref.~\cite{Poulin:2018cxd} or in the $\alpha$-attractor EDE model of Ref.~\cite{Braglia:2020bym}, provide a better fit to Planck polarization data and lead to an even larger $H_0$.
It would be interesting to explore different choices of the potential in the EMG framework.

\section*{Acknowledgments}

The authors thank  Guillermo F. Abellan, Riccardo Murgia and Vivian Poulin for help with the implementation of IMINUIT in MontePython and for comments on the draft. MBr thanks Xingang Chen for pointing out the possibility that the field $\sigma$ climbs up the potential at very early times due to quantum fluctuations.  
MBa and FF acknowledge financial contribution from the contract ASI/INAF for the Euclid mission n.2018-23-HH.0.
FF acknowledges financial support by ASI Grant 2020-9-H.0. 
KK received funding from the European Research Council under the European Union’s Horizon 2020 research and innovation programme (grant agreement No. 646702 "CosTesGrav"). KK is also supported by the UK STFC ST/S000550/1. Numerical computations for this research were done on the Sciama High Performance Compute cluster, which is supported by the ICG, SEPNet, and the University of Portsmouth.

    \newpage

\appendix
\section{Tables} \label{sec:appendix}

\begin{table*}[h!]
	{\small
		\centering
		\begin{tabular}{|l||c|c|c|}
			\hline
			\hline 
			& $\Lambda$CDM & EDE & EMG \\
			\hline
			$10^{2}\omega_{\rm b}$                        &  $ 2.256\pm 0.013\,$ $(2.255)$        &  $2.280\pm 0.018\,$ $(2.286)$  &   $2.273\pm 0.017 \,$ $(2.281)$ \\
			$\omega_{\rm c}$                        & $0.1182\pm 0.0009\,$ $(0.1184)$                    &$0.1253^{+0.0033}_{-0.0038}\,$ $(0.1242)$     &$ 0.1282^{+0.0042}_{-0.0033}\,$ $(0.1302)$ \\
			$100*\theta_{s }$             &$1.04209\pm 0.00028\,$ $(1.04216)$   &  $1.04152\pm 0.00036,\,$ $(1.04170)$  & $1.04118^{+0.00040}_{-0.00046}\,$ $(1.04120)$  \\
			$\tau_\textup{reio }$                               & $0.058\pm 0.007\,$ $(0.052)$  & $0.058^{+0.007}_{-0.008}\,$ $(0.059)$ & $ 0.056\pm 0.007\,$ $(0.057)$ \\
			$\ln \left(  10^{10} A_{\rm s} \right)$ & $3.049\pm 0.014\,$ $(3.038)$     &  $3.059\pm 0.016\,$ $(3.059)$  &$3.061\pm 0.015\,$ $(3.067)$ \\
			$n_{\rm s}$                             & $0.9701\pm 0.0036\,$ $(0.9710)$  &  $ 0.9783^{+0.0054}_{-0.0061}\,$ $(0.9813)$  &$0.9782\pm 0.0055\,$ $(0.9849)$ \\
			$\sigma_i$ [M$_\mathrm{pl}$]                       &  $-$   &  $< 0.70\,$ $(0.48)$  &  $0.49^{+0.11}_{-0.06}\,$ $(0.53)$   \\
			$V_0$                        &  $-$    &  $2.21^{+0.07}_{-0.38}\,$ $(2.09)$  &  $2.21^{+0.10}_{-0.15}\,$ $(2.25)$  \\
			$\xi$                        &  $-$    &  $-$  &  $< 0.42\,$ $(0.18)$ \\
			\hline
			$H_0$ [km s$^{-1}$Mpc$^{-1}$]             &  $68.82\pm 0.39\,$ $(68.74)$      &$70.57^{+0.77}_{-0.98}\,$ $(70.90)$  & $71.00^{+0.87}_{-0.79}\,$ $(71.59)$  \\
			$r_s$ [Mpc]                             &  $147.37\pm 0.22 \,$ $(147.33)$ &  $143.5\pm 1.8\,$ $(143.78)$ &      $142.2^{+1.5}_{-2.0}\,$ $(141.21)$ \\
			$\sigma_8$                              &  $0.821\pm 0.006 \,$ $(0.818)$ &  $0.832^{+0.009}_{-0.011}\,$ $(0.831)$&$  0.830\pm 0.008\,$ $(0.850)$  \\
			$S_8$                              &  $0.817\pm 0.010\,$ $(0.815)$ &  $0.829\pm 0.013\,$ $(0.820)$ &$ 0.829\pm 0.011\,$ $(0.847)$ \\			$\log_{10}\,z_c$                        &  $-$   &  $3.58^{+0.04}_{-0.16}\,$ $(3.53)$ &$3.60^{+0.06}_{-0.05}\,$ $(3.63)$\\
			$f_{\rm scf}$                        &  $-$  &    $< 0.119\,$ $(0.057)$ &$0.084^{+0.030}_{-0.021}\,$ $(0.099)$ \\
			$\xi\sigma^2_i$ [$M_\mathrm{pl}^2$]              & $-$      &$-$   & $< 0.067\, $ $(0.050)$  \\
			$\gamma_{\rm PN}-1$                           &$-$   &  $-$ & $> -1.7\cdot 10^{-9}\,$ $(-8.9\cdot 10^{-9})$\\
			\hline
			$\Delta \chi^2$                         & $-$ & -9.3 & -16.0 \\
			$\ln B_{ij}$                         & $-$ & $+1.5$ &$+1.0$\\
			\hline
			\hline
	\end{tabular}}
	\break
	\break{\break}
		{\small
		\centering
		\begin{tabular}{|l||c|c|c|}
			\hline
			\hline P18 + BAO + FS + SN + $H_0$
			& $\Lambda$CDM & EDE & EMG \\
			\hline
			Planck high-$\ell$ TTTEEE                     & 2350.07  & 2352.08 &  2347.75\\
			Planck low-$\ell$ EE                   & 395.70 & 396.69 & 396.37 \\
			Planck low-$\ell$ TT             & 22.32 & 21.51 & 21.52 \\
			Planck lensing                               & 9.37 & 9.36 & 9.17\\
			BAO BOSS low-$z$ & 2.21 & 2.74 & 2.06\\
			BAO DR12 FS + BAO, high-$z$ NGC                             & 65.13  & 65.15 &  67.64\\
			BAO DR12 FS + BAO, high-$z$ SGC                       &  62.63 & 63.29 & 62.83\\
			BAO DR12 FS + BAO, low-$z$ NGC                    & 70.06 & 70.53 & 69.89\\
			Pantheon       & 1026.86 & 1026.93 & 1026.88\\
			$H_0$              & 18.57 &  5.35 &  2.81\\
			\hline
			Total                         & 4022.94 & 4013.64 & 4006.92 \\
			\hline
			\hline
	\end{tabular}}
	\break
	\break
	\caption{\label{tab:P18BAOFSSNH0} 
		[Upper table] Constraints on main and derived parameters  considering 
		the data set {\bf P18 + BAO + FS + SN + $H_0$}  for $\Lambda$CDM, $\xi=0$ and $\xi\geq0$.   We report mean values and the 68\% CL, except for the case of upper or lower limits, for which we report the 95\% CL. We also report the best-fit values in round brackets. [Lower table] 		Best-fit $\chi^2$ per experiment for 
		the data set {\bf P18 + BAO + FS + SN + $H_0$}  for $\Lambda$CDM, EDE and EMG model.}
\end{table*}

\begin{table*}[h!]
	{\small
		\centering
		\begin{tabular}{|l||c|c|c|}
			\hline
			\hline 
			& $\Lambda$CDM & EDE & EMG \\
			\hline
			$10^{2}\omega_{\rm b}$                      &$2.261\pm 0.014\,$ $(2.263)$ &$2.283\pm 0.018\,$ $(2.292)$ & $2.275\pm 0.018\,$ $(2.284)$\\
			$\omega_{\rm c}$                       &$0.1175\pm 0.0011\,$ $(0.1170)$ & $0.1253^{+0.0036}_{-0.0044}\,$ $(0.1285)$ & $0.1288\pm 0.0046\,$ $( 0.131)$\\
			$100*\theta_{s }$    &   $1.04216\pm 0.00029\,$ $(1.04200)$    &  $1.04153\pm 0.00038\,$ $(1.04135)$ & $1.04114\pm 0.00048\,$ $(1.04107)$ \\
			$\tau_\textup{reio }$                &$0.061^{+0.007}_{-0.008}\,$ $(0.060)$   & $0.060^{+0.007}_{-0.008}\,$ $(0.061)$ &   $0.058^{+0.007}_{-0.008}\,$ $(0.056)$          \\
			$\ln \left(  10^{10} A_{\rm s} \right)$ &$3.053^{+0.014}_{-0.016} \,$ $(3.050)$ &$3.062\pm 0.016\,$  $(3.072)$& $3.067\pm 0.016\,$ $(3.067)$\\
			$n_{\rm s}$                     & $0.9707\pm 0.0040\,$ $(0.9733)$ & $0.9788\pm 0.0061$\, $(0.9849)$ &         $0.9800\pm 0.0059\,$ $(0.9870)$\\
			$\sigma_i$ [M$_\mathrm{pl}$]     &  $-$                 & $0.48^{+0.14}_{-0.09}\,$ $(0.58)$ &    $ 0.50^{+0.12}_{-0.07} \,$ $(0.56)$\\
			$V_0$                       & $-$  &$2.23^{+0.10}_{-0.45}\,$ $(1.97)$ &$2.22^{+0.11}_{-0.13}\,$ $(2.24)$  \\
			$\xi$                        &$-$ & $-$ & $< 0.39\,$ $(0.17)$\\
			\hline
			$H_0$ [km s$^{-1}$Mpc$^{-1}$]            &$69.13\pm 0.49\,$ $(69.25)$ & $70.85\pm0.92\,$ $(71.38)$ & $ 71.21\pm 0.93 \,$ $(71.87)$ \\
			$r_s$ [Mpc]                             &$147.49\pm 0.25\,$ $(147.61)$ & $143.4\pm 1.9\,$ $(141.83)$ & $141.9^{+1.9}_{-2.2} \,$ $(140.70)$ \\
			$\sigma_8$                              &$0.820\pm 0.006\,$ $(0.818)$ & $0.833\pm 0.011\,$ $(0.842)$ & $0.833\pm 0.008\,$ $(0.836)$  \\
			$S_8$                              &$0.811\pm 0.011\,$ $(0.806)$ & $0.827\pm 0.016\,$ $(0.838)$ & $0.831\pm 0.014\,$ $(0.833)$ \\	
			$\log_{10}\,z_c$                        & $-$ & $3.59^{+0.06}_{-0.19}\,$ $(3.50)$ & $3.60^{+0.06}_{-0.04}\,$ $(3.64)$\\
			$f_{\rm scf}$                       &  $-$& $< 0.134\,$ $(0.083)$ & $ 0.088^{+0.033}_{-0.025} \,$ $(0.107)$ \\
			$\xi\sigma^2_i$ [$M_\mathrm{pl}^2$]              &  $-$& $-$ & $< 0.072\,$ $(0.053)$ \\
			$\gamma_{\rm PN}-1$                           & $-$ & $-$ & $> -1.7\cdot 10^{-9}\,$ $(-1.8\cdot10^{-9})$ \\
			\hline
			$\Delta \chi^2$                         & $-$ & $-11.5$ &$-17.1$\\
			$\ln B_{ij}$                         & $-$ & $+1.8$ &$-0.2$\\
			\hline
			\hline
	\end{tabular}}
	\break
	\break{\break}
		{\small
		\centering
		\begin{tabular}{|l||c|c|c|}
			\hline
			\hline  P18 + SN + $H_0$
			& $\Lambda$CDM & EDE & EMG \\
			\hline
			Planck high-$\ell$ TTTEEE                     &2351.75 & 2352.22 &2349.25\\
			Planck low-$\ell$ EE                   &396.94 & 397.51 & 396.23\\
			Planck low-$\ell$ TT             &22.08 & 21.41 & 21.29\\
			Planck lensing                              &9.59 & 9.07 &9.32\\
			Pantheon       & 1026.96& 1026.87 &1026.86\\
			$H_0$              &14.76 & 3.50 & 2.00\\
			\hline
			Total                         & 3822.08& 3810.58 & 3804.97\\
			\hline
			\hline
	\end{tabular}}
	\break
	\break
	\caption{\label{tab:P18SNH0} 
		[Upper table] Constraints on main and derived parameters  considering 
		the data set {\bf P18 + SN + $H_0$}  for $\Lambda$CDM, $\xi=0$ and $\xi\geq0$.   We report mean values and the 68\% CL, except for the case of upper or lower limits, for which we report the 95\% CL. We also report the best-fit values in round brackets. [Lower table] 		Best-fit $\chi^2$ per experiment for 
		the data set  {\bf P18 + SN + $H_0$}  for $\Lambda$CDM, EDE and EMG model.}
\end{table*}

\begin{table*}[h!]
	{\small
		\centering
		\begin{tabular}{|l||c|c|c|}
			\hline
			\hline 
			& $\Lambda$CDM & EDE & EMG \\
			\hline
			$10^{2}\omega_{\rm b}$                      &$2.262\pm 0.013\,$ $(2.265)$ & $ 2.277\pm 0.016\,$ $(2.276)$ & $2.272\pm 0.016\,$ $(2.275)$\\
			$\omega_{\rm c}$                       &$0.1174\pm 0.0008\,$ $(0.1178)$ & $0.1218^{+0.0022}_{-0.0034}\,$ $(0.1228)$ & $0.1234^{+0.0028}_{-0.0047}\,$ $(0.1262)$\\
			$100*\theta_{s }$           & $1.04213\pm 0.00029\,$ $(1.04229)$ &$1.04160^{+0.00052}_{-0.00034}\,$ $()$ & $1.04154\pm 0.00043\,$ $(1.04148)$\\
			$\tau_\textup{reio }$                &$ 0.055\pm 0.007\,$ $(0.057)$ & $0.055\pm 0.007\,$ $(0.058)$ & $0.054\pm 0.007\,$ $(0.057)$             \\
			$\ln \left(  10^{10} A_{\rm s} \right)$ & $3.041\pm 0.014\,$ $(3.047)$ & $3.044\pm 0.015\,$ $(3.042)$ & $3.047\pm 0.015\,$ $(3.058)$\\
			$n_{\rm s}$               & $0.9716\pm 0.0035\,$ $(0.9719)$ & $0.9756^{+0.0043}_{-0.0053}\,$ $(0.9752)$ & $ 0.9755^{+0.0046}_{-0.0054}\,$ $(0.9791)$              \\
			$\sigma_i$ [M$_\mathrm{pl}$]                      & $-$& $<0.60\,$ $(0.47)$ &  $0.39^{+0.15}_{-0.10}\,$ $(0.50)$  \\
			$V_0$                       & $-$& $2.59^{+0.72}_{-0.64}\,$ $(3.21)$ &$2.44^{+0.76}_{-0.50}\,$ $(2.05)$  \\
			$\xi$                        & $-$&$-$ & $<0.63\,$ $(0.14)$\\
			\hline
			$H_0$ [km s$^{-1}$Mpc$^{-1}$]            &$69.17\pm 0.35\,$ $(69.09)$ & $70.40\pm 0.76\,$ $(70.75)$ & $70.63^{+0.80}_{-1.0}\,$ $(71.59)$ \\
			$r_s$ [Mpc]                             &$147.51\pm 0.21\,$ $(147.38)$ &$ 145.0^{+1.7}_{-1.3}\,$ $(144.43)$ &$144.3^{+2.3}_{-1.5}\,$ $(142.75)$ \\
			$\sigma_8$                              &$0.815\pm 0.005\,$ $(0.819)$ &$0.819^{+0.006}_{-0.008}\,$ $(0.81682)$ & $0.819^{+0.006}_{-0.007}\,$ $(0.820)$\\
			$S_8$                              &$0.805\pm 0.008\,$ $(0.811)$  & $0.808\pm 0.010\,$ $(0.804)$ & $0.809\pm 0.009\,$ $(0.807)$\\	
			$\log_{10}\,z_c$                        &$-$ & $3.72^{+0.37}_{-0.26} \,$ $(4.12)$ & $3.66^{+0.04}_{-0.20}\,$ $( 3.52)$ \\
			$f_{\rm scf}$                       & $-$& $<0.101\,$ $(0.064)$ & $<0.121\,$ $(0.085)$\\
			$\xi\sigma^2_i$ [$M_\mathrm{pl}^2$]              & $-$&$-$ & $<0.054\,$ $(0.030)$ \\
			$\gamma_{\rm PN}-1$                           &$-$ & $-$& $> -1.8\cdot 10^{-9}\,$ $(-8.0\cdot 10^{-10})$\\
			\hline
			$\Delta \chi^2$                         &$-$ & $-11.0$ &$-11.5$\\
			$\ln B_{ij}$                         & $-$ & $-0.4$ &$-0.12$\\
			\hline
			\hline
	\end{tabular}}
	\break
	\break{\break}
		{\small
		\centering
		\begin{tabular}{|l||c|c|c|}
			\hline
			\hline  P18 + BAO + FS + SN + $H_0$ + $S_8$
			& $\Lambda$CDM & EDE & EMG \\
			\hline
			Planck high-$\ell$ TTTEEE                     &2351.17 &2351.13 &2351.51\\
			Planck low-$\ell$ EE                   &396.43 & 396.47 & 396.48 \\
			Planck low-$\ell$ TT             &22.36 &21.70 & 22.19\\
			Planck lensing                              &9.32 & 10.09&10.46\\
			BAO BOSS low-$z$ & 2.65&2.96 & 2.91\\
			BAO DR12 FS + BAO, high-$z$ NGC                             & 64.76& 64.08& 65.53\\
			BAO DR12 FS + BAO, high-$z$ SGC                      &63.11 & 63.23& 63.00\\
			BAO DR12 FS + BAO, low-$z$ NGC & 70.57 & 71.14& 70.54\\
			Pantheon       &1026.89 & 1026.97 &1026.98\\
			$H_0$              & 15.88 & 6.00& 2.90\\
			$S_8$              &5.66 & 4.02 &4.82 \\
			\hline
			Total                         & 4028.81& 4017.81 & 4017.35\\
			\hline
			\hline
	\end{tabular}}
	\break
	\break
	\caption{\label{tab:P18BAOFSSNH0S8} 
		[Upper table] Constraints on main and derived parameters  considering 
		the data set {\bf P18 + BAO + FS + SN + $H_0$ + $S_8$}  for $\Lambda$CDM, $\xi=0$ and $\xi\geq0$.   We report mean values and the 68\% CL, except for the case of upper or lower limits, for which we report the 95\% CL. We also report the best-fit values in round brackets. [Lower table] 		Best-fit $\chi^2$ per experiment for 
		the data set  {\bf P18 + BAO + FS + SN + $H_0$ + $S_8$}  for $\Lambda$CDM, EDE and EMG model.}
\end{table*}

\begin{table*}[h!]
	{\small
		\centering
		\begin{tabular}{|l||c|c|c|}
			\hline
			\hline
			& $\Lambda$CDM & EDE & EMG \\
			\hline
			$10^{2}\omega_{\rm b}$                        &  $ 2.243\pm 0.013\,$ $(2.251)$        &  $2.245^{+0.015}_{-0.016}\,$ $(2.240)$  &  $2.244\pm 0.015\,$ $(2.247)$ \\
			$\omega_{\rm c}$                        & $0.1195\pm 0.0009\,$ $(0.1186)$                    &$0.1206^{+0.0008}_{-0.0019}\,$ $(0.1200)$     &$0.1206^{+0.0011}_{-0.0019}\,$ $(0.1234)$\\
			$100*\theta_{s }$             &$1.04193\pm 0.00029\,$ $(1.04199)$   &  $1.04182\pm 0.00032,\,$ $(1.04181)$  & $1.04181^{+0.00033}_{-0.00029}\,$ $(1.04168)$ \\
			$\tau_\textup{reio }$                               & $0.054\pm 0.007\,$ $(0.059)$  & $0.054\pm 0.007,\,$ $(0.054)$ &$0.054\pm 0.007\,$ $(0.54)$\\
			$\ln \left(  10^{10} A_{\rm s} \right)$ & $3.043\pm 0.014 \,$ $(3.050)$     &  $3.045\pm 0.014\,$ $(3.044)$  &$3.045\pm 0.014\,$ $(3.0491)$\\
			$n_{\rm s}$                             & $0.9666\pm 0.0037\,$ $(0.9699)$  &  $ 0.9678^{+0.0037}_{-0.0047}\,$ $(0.9663)$  &$0.9673\pm 0.0044\,$ $(0.9686)$\\
			$\sigma_i$ [M$_\mathrm{pl}$]                       &  $-$   &  $< 0.50\,$ $(0.05)$  &$< 0.45\,$ $(0.31)$   \\
			$V_0$                        &  $-$    &  $2.14\pm 0.78\,$ $(0.69)$& $2.47^{+0.86}_{-0.39}\,$ $(2.28)$\\
			$\xi$                        &  $-$    &  $-$  &$<0.81\,$ $(0.18)$ \\
			\hline
			$H_0$ [km s$^{-1}$Mpc$^{-1}$]             &  $68.16\pm 0.41\,$ $(68.55)$      &$68.46^{+0.42}_{-0.68}\,$ $(67.90)$  &$68.39^{+0.50}_{-0.67}\,$  $(68.94)$\\
			$r_s$ [Mpc]                             &  $147.16\pm 0.22 \,$ $(147.32)$ &  $146.53^{+0.94}_{-0.23}\,$ $(147.08)$ &$146.59^{+0.90}_{-0.38}\,$    $(145.17)$  \\
			$\sigma_8$                              &  $0.822\pm 0.0058 \,$ $(0.823)$ &  $0.823^{+0.006}_{-0.007}\,$ $(0.824)$&$0.822\pm 0.007\,$ $(0.824)$\\
			$S_8$                              &  $0.830\pm 0.010\,$ $(0.823)$ &  $0.831\pm 0.011\,$ $(0.836)$ &$0.830\pm 0.011$  $(0.834)$ \\			$\log_{10}\,z_c$                        &  $-$   &  $3.26^{+0.65}_{-0.72}\,$ $(2.07)$ & $3.44^{+0.52}_{-0.17} \,$ $(3.54)$\\
			$f_{\rm scf}$                        &  $-$  &    $< 0.0617\,$ $( 0.0004)$ &$<0.0726\,$ $(0.037)$\\
			$\xi\sigma^2_i$ [$M_\mathrm{pl}^2$]              & $-$      &$-$   &$<0.0381\,$ $(0.0172)$  \\
			$\gamma_{\rm PN}-1$                           &$-$   &  $-$ &$> -1.7\cdot 10^{-8}\,$  $(-5.0\cdot 10^{-10})$ \\
			\hline
			$\Delta \chi^2$                         & $-$ & $-1.2$ & $-2.6$ \\
			$\ln B_{ij}$                         & $-$ & $-1.3$ &$-2.7$\\
			\hline
			\hline
	\end{tabular}}
	\break
	\break{\break}
		{\small
		\centering
		\begin{tabular}{|l||c|c|c|}
			\hline
			\hline P18 + BAO + FS+ SN 
			& $\Lambda$CDM & EDE & EMG \\
			\hline
			Planck high-$\ell$ TTTEEE                     & 2347.99  & 2346.77& 2345.32 \\
			Planck low-$\ell$ EE                   &396.89& 396.00 & 396.04\\
			Planck low-$\ell$ TT             &22.69& 23.23&   23.34\\
			Planck lensing                               & 8.82& 8.86& 8.80\\
			BAO BOSS low-$z$ &  2.00& 1.33 & 1.44\\
			BAO DR12 FS + BAO, high-$z$ NGC                             &  65.78 & 67.86& 67.91 \\
			BAO DR12 FS + BAO, high-$z$ SGC                       & 62.42 & 61.76& 61.69 \\
			BAO DR12 FS + BAO, low-$z$ NGC                    & 69.82 & 69.25 & 69.17\\
			Pantheon       &1026.89 & 1027.09& 1027.02\\
			\hline
			Total                         & 4003.30 & 4002.15 &4000.74 \\
			\hline
			\hline
	\end{tabular}}
	\break
	\break
	\caption{\label{tab:P18BAOFSSN} 
		[Upper table] Constraints on main and derived parameters  considering 
		the data set {\bf P18 + BAO + FS + SN }  for $\Lambda$CDM, $\xi=0$ and $\xi\geq0$.   We report mean values and the 68\% CL, except for the case of upper or lower limits, for which we report the 95\% CL. We also report the best-fit values in round brackets. [Lower table] 		Best-fit $\chi^2$ per experiment for 
		the data set {\bf P18 + BAO + FS + SN}  for $\Lambda$CDM, EDE and EMG model.}
\end{table*}

    \newpage



\end{document}